\documentclass[12pt]{article}
\usepackage{amssymb}
\usepackage{graphicx}
\begin{document}
\newcommand{\beq}{\begin{equation}}
\newcommand{\eeq}{\end{equation}}
\newcommand{\beqa}{\begin{eqnarray}}
\newcommand{\eeqa}{\end{eqnarray}}
\newcommand{\beqar}{\begin{eqnarray*}}
\newcommand{\eeqar}{\end{eqnarray*}}
\newcommand{\al}{\alpha}
\newcommand{\be}{\beta}
\newcommand{\del}{\delta}
\newcommand{\D}{\Delta}
\newcommand{\eps}{\epsilon}
\newcommand{\ga}{\gamma}
\newcommand{\Ga}{\Gamma}
\newcommand{\ka}{\kappa}
\newcommand{\nn}{\nonumber}
\newcommand{\inn}{\!\cdot\!}
\newcommand{\h}{\eta}
\newcommand{\ii}{\iota}
\newcommand{\kk}{\varphi}
\newcommand\F{{}_3F_2}
\newcommand{\la}{\lambda}
\newcommand{\La}{\Lambda}
\newcommand{\na}{\prt}
\newcommand{\Om}{\Omega}
\newcommand{\om}{\omega}
\newcommand{\p}{\phi}
\newcommand{\sig}{\sigma}
\renewcommand{\t}{\theta}
\newcommand{\z}{\zeta}
\newcommand{\ssc}{\scriptscriptstyle}
\newcommand{\eg}{{\it e.g.,}\ }
\newcommand{\ie}{{\it i.e.,}\ }
\newcommand{\labell}[1]{\label{#1}} 
\newcommand{\reef}[1]{(\ref{#1})}
\newcommand\prt{\partial}
\newcommand\veps{\varepsilon}
\newcommand{\pol}{\varepsilon}
\newcommand\vp{\varphi}
\newcommand\dS{\dot{\cal S}}
\newcommand\dH{\dot{\cal H}}
\newcommand\dcB{\dot{\cal B}}
\newcommand\dA{\dot {A}}
\newcommand\dB{\dot {B}}
\newcommand\dG{\dot {G}}
\newcommand\dP{\dot {\phi}}
\newcommand\ls{\ell_s}
\newcommand\cF{{\cal F}}
\newcommand\cA{{\cal A}}
\newcommand\cS{{\cal S}}
\newcommand\cT{{\cal T}}
\newcommand\cV{{\cal V}}
\newcommand\cL{{\cal L}}
\newcommand\cM{{\cal M}}
\newcommand\cN{{\cal N}}
\newcommand\cG{{\cal G}}
\newcommand\cH{{\cal H}}
\newcommand\cI{{\cal I}}
\newcommand\cJ{{\cal J}}
\newcommand\cl{{\iota}}
\newcommand\cP{{\cal P}}
\newcommand\cQ{{\cal Q}}
\newcommand\cg{{\it g}}
\newcommand\cR{{\cal R}}
\newcommand\cB{{\cal B}}
\newcommand\cO{{\cal O}}
\newcommand\tcO{{\tilde {{\cal O}}}}
\newcommand\bg{\bar{g}}
\newcommand\bb{\bar{b}}
\newcommand\bH{\bar{H}}
\newcommand\bF{\bar{F}}
\newcommand\bX{\bar{X}}
\newcommand\bK{\bar{K}}
\newcommand\bA{\bar{A}}
\newcommand\bZ{\bar{Z}}
\newcommand\bxi{\bar{\xi}}
\newcommand\bphi{\bar{\phi}}
\newcommand\bpsi{\bar{\psi}}
\newcommand\bprt{\bar{\prt}}
\newcommand\bet{\bar{\eta}}
\newcommand\btau{\bar{\tau}}
\newcommand\hF{\hat{F}}
\newcommand\hA{\hat{A}}
\newcommand\hT{\hat{T}}
\newcommand\htau{\hat{\tau}}
\newcommand\hD{\hat{D}}
\newcommand\hf{\hat{f}}
\newcommand\hg{\hat{g}}
\newcommand\hp{\hat{\phi}}
\newcommand\hi{\hat{i}}
\newcommand\ha{\hat{a}}
\newcommand\hb{\hat{b}}
\newcommand\hQ{\hat{Q}}
\newcommand\hP{\hat{\Phi}}
\newcommand\hS{\hat{S}}
\newcommand\hX{\hat{X}}
\newcommand\tL{\tilde{\cal L}}
\newcommand\hL{\hat{\cal L}}
\newcommand\tG{{\widetilde G}}
\newcommand\tg{{\widetilde g}}
\newcommand\tphi{{\widetilde \phi}}
\newcommand\tPhi{{\widetilde \Phi}}
\newcommand\te{{\tilde e}}
\newcommand\tk{{\tilde k}}
\newcommand\tf{{\tilde f}}
\newcommand\ta{{\tilde a}}
\newcommand\tb{{\tilde b}}
\newcommand\tR{{\tilde R}}
\newcommand\teta{{\tilde \eta}}
\newcommand\tF{{\widetilde F}}
\newcommand\tK{{\widetilde K}}
\newcommand\tE{{\widetilde E}}
\newcommand\tpsi{{\tilde \psi}}
\newcommand\tX{{\widetilde X}}
\newcommand\tD{{\widetilde D}}
\newcommand\tO{{\widetilde O}}
\newcommand\tS{{\tilde S}}
\newcommand\tB{{\widetilde B}}
\newcommand\tA{{\widetilde A}}
\newcommand\tT{{\widetilde T}}
\newcommand\tC{{\widetilde C}}
\newcommand\tV{{\widetilde V}}
\newcommand\thF{{\widetilde {\hat {F}}}}
\newcommand\Tr{{\rm Tr}}
\newcommand\tr{{\rm tr}}
\newcommand\STr{{\rm STr}}
\newcommand\hR{\hat{R}}
\newcommand\MZ{\mathbb{Z}}
\newcommand\MR{\mathbb{R}}
\newcommand\M[2]{M^{#1}{}_{#2}}

\newcommand\bS{\textbf{ S}}
\newcommand\bI{\textbf{ I}}
\newcommand\bJ{\textbf{ J}}

\begin{titlepage}
\begin{center}

\vskip 2 cm
{\LARGE \bf    $O(9,25)$ symmetry of heterotic string theory  \\  \vskip 0.25 cm   at orders $\alpha'$, $\alpha'^2$
 }\\
\vskip 1.25 cm
 Mohammad R. Garousi\footnote{garousi@um.ac.ir}
 
\vskip 1 cm
{{\it Department of Physics, Faculty of Science, Ferdowsi University of Mashhad\\}{\it P.O. Box 1436, Mashhad, Iran}\\}
\vskip .1 cm
 \end{center}
\begin{abstract}

In a recent study, we have observed that by imposing a truncated T-duality transformation on the circular reduction of the bosonic couplings in the heterotic theory at four- and six-derivative orders, we can calculate these couplings in a particular YM gauge where the YM potential vanishes but its field strength remains non-zero. Importantly, the coupling constants are independent of the gauge choice, so these results are valid across different YM gauge choices.

In this work, we explore the cosmological reduction of these couplings when the YM gauge fields belong to the Cartan subalgebra of $SO(32)$ or $E_8 \times E_8$. We demonstrate that after applying appropriate one-dimensional field redefinitions and total derivative terms, the couplings can be expressed in a proposed $O(9,25)$-invariant canonical form, which is the extension of the canonical $O(9,9)$-invariant form for just the NS-NS fields proposed by Hohm and Zwiebach. This $O(9,25)$-invariant expression is in terms of the trace of the first time derivative of the generalized metric, which encompasses both the YM field and the NS-NS fields.

\end{abstract}
\end{titlepage}

\section{Introduction} \label{intro}

Recent work has leveraged the background independence and Yang-Mills (YM) gauge invariance of the effective action in heterotic string theory to systematically derive the bosonic four- and six-derivative couplings \cite{Garousi:2024avb,Garousi:2024imy}. Interestingly, it was also shown that all odd-derivative couplings vanish in the heterotic string theory \cite{Garousi:2024vbz}.
The key insight behind this progress was to choose a specific YM gauge where the YM potential is zero but its field strength remains non-zero. In this gauge, all commutator terms in the YM field strength and its derivatives conveniently vanish, simplifying the identification of the basis at each derivative order \cite{Garousi:2024avb,Garousi:2024imy,Garousi:2024vbz}. The coupling constant of each coupling in the basis is yet arbitrary.
The background independence was then leveraged to consider a particular background with a circular dimension and perform a circular reduction, which induces a non-geometric $O(1,1,\mathbb{Z})$ symmetry in the effective action. A truncated form of this emergent $\mathbb{Z}_2$ symmetry has then been used as a powerful tool to determine the coupling constants in the heterotic string theory.
In this paper, we aim to further verify these couplings by showing that the cosmological reduction of these couplings indeed satisfies the expected $O(9,25)$-symmetry \cite{Sen:1991zi,Hassan:1991mq,Hohm:2014sxa}.

The cosmological reduction of  Neveu-Schwarz-Neveu-Schwarz (NS-NS) couplings of  the classical effective action at any order of $\alpha'$ should exhibit $O(d,d,\MR)$ symmetry \cite{Sen:1991zi, Hohm:2014sxa}. It has been shown in 
 \cite{Hohm:2015doa, Hohm:2019jgu} that  cosmological action at order $\alpha'$ and higher   can be expressed in terms of  only the first time-derivative of the $O(d,d)$-valued generalized metric ${\cal S}$.  The trace of an odd number of  $\dot{\cS}$ vanishes \cite{Hohm:2019jgu}, and the one-dimensional action can be expressed in a canonical form as the following $O(d,d)$-invariant expansion \cite{Hohm:2015doa, Hohm:2019jgu}:
\beqa
S_{\rm eff}^c&=&S_0^c-\frac{2}{\kappa^2}\int dt e^{-\phi}\bigg(\alpha' c_{2,0}\tr(\dS^4)+\alpha'^2c_{3,0}\tr(\dS^6)\nn\\
&&\qquad\quad\qquad\quad+\alpha'^3[c_{4,0}\tr(\dS^8)+c_{4,1}(\tr(\dS^4))^2]\nn\\
&&\qquad\quad\qquad\quad+\alpha'^4[c_{5,0}\tr(\dS^{10})+c_{5,1}\tr(\dS^6)\tr(\dS^4)]
+\cdots\bigg)\,.\labell{cosm}
\eeqa
In the above action, the coefficients $c_{m,n}$ depend on the specific theory. By examining the cosmological reduction of only the pure gravity components of the couplings in various theories, one can determine the coefficients $c_{m,n}$. These coefficients up to order $\alpha'^3$ have been determined in \cite{Codina:2021cxh}. 
It has been shown in \cite{Garousi:2021ikb,Garousi:2021ocs,Gholian:2023kjj} that the cosmological reduction of the NS-NS couplings at orders $\alpha'^2$ and $\alpha'^3$, which were found in \cite{Garousi:2019mca,Garousi:2023kxw,Garousi:2020gio} by T-duality of circular reduction, satisfy the above canonical form after using appropriate one-dimensional field redefinitions and integration by parts.

It has been established in the literature \cite{Hohm:2014sxa} that when the effective action is extended to include YM fields belonging to the full $SO(32)$ or $E_8 \times E_8$ algebra, the true symmetry of the torus $T^{d}$ reduction of the heterotic effective action is the $O(d,d)$ symmetry, and the NS-NS and YM scalars in the base space appear in an $O(d,d)$-valued generalized metric and vector.
However, if the gauge fields are restricted to the Cartan subalgebra of $SO(32)$ or $E_8 \times E_8$, where the YM fields commute, then the symmetry increases to $O(d,d+16)$. In this case, the NS-NS and YM fields can be combined into an $O(d,d+16)$-valued generalized metric, which encapsulates the symmetry between the NS-NS and YM sectors \cite{Hohm:2014sxa}.
In this restricted case, the cosmological action should take the same canonical form as for the NS-NS couplings \reef{cosm}, but the generalized metric should now incorporate contributions from both the NS-NS fields and the YM fields. We expect the one-dimensional couplings to appear in the following form:
\beqa
S_{\rm eff}^c&=&S_0^c-\frac{2}{\kappa^2}\int dt e^{-\phi}\bigg(\alpha' c_{2,0}\tr(\dH^4)+\alpha'^2c_{3,0}\tr(\dH^6)\nn\\
&&\qquad\quad\qquad\quad+\alpha'^3[c_{4,0}\tr(\dH^8)+c_{4,1}(\tr(\dH^4))^2]\nn\\
&&\qquad\quad\qquad\quad+\alpha'^4[c_{5,0}\tr(\dH^{10})+c_{5,1}\tr(\dH^6)\tr(\dH^4)]
+\cdots\bigg)\,,\labell{cosmH}
\eeqa
where the $O(9,25)$-valued  matrix $\mathcal{H}$  includes both the NS-NS and YM fields \cite{Maharana:1992my,Hohm:2014sxa}. 
When the YM fields are zero, the matrix $\mathcal{H}$ reduces to the matrix $\mathcal{S}$.
Hence, the coefficients $c_{m,n}$ should be  the same as the coefficients in \reef{cosm}.

It has been observed that the YM fields appear in the 10-dimensional couplings at four-, six-derivative and presumably all higher orders, only through multiples of $\Tr(FF)$ and their derivatives \cite{Garousi:2024avb,Garousi:2024imy}. Hence, restricting the YM field to the Cartan subalgebra does not remove any YM gauge-invariant coupling from the couplings found in \cite{Garousi:2024avb,Garousi:2024imy}.
Consequently, to validate these couplings  by verifying their cosmological symmetry, it is sufficient to consider the YM gauge field belonging to the Cartan subalgebra of $SO(32)$ or $E_8 \times E_8$ and show that they have the $O(9,25)$ symmetry and satisfy the canonical form \reef{cosmH}.
There are no couplings in the cosmological action \reef{cosmH} that involve an odd number of derivatives. This immediately verifies the observation that there are no odd-derivative YM gauge-invariant couplings in heterotic theory \cite{Garousi:2024vbz}.

If the classical effective action  $\!\!\bS_{\rm eff}$, its corresponding cosmological reduction $\!\!\bS^c_{\rm eff}$, and the $O(9,25)$-invariant cosmological action $S_{\rm eff}^c$ have the following $\alpha'$-expansions:
\beqa
\bS_{\rm eff} = \sum^\infty_{m=0}\alpha'^m\bS_m &,&\bS^c_{\rm eff}(\psi) = \sum^\infty_{m=0}\alpha'^m\bS^c_m(\psi)\,\,,\,\, S^c_{\rm eff}(\psi)=\sum^\infty_{m=0}\alpha'^m S^c_m(\psi), \labell{seff}
\eeqa
where $\psi$ collectively represents the one-dimensional functions, then the relation between the $S_m^c$'s and the $\bS_m^c$'s up to order $\alpha'^2$ is given by \cite{Garousi:2021ocs}:
\beqa
S_0^c(\psi)&=&\!\!\bS_0^c(\psi)+\cJ_0,\nn\\
S_1^c(\psi)&=&\!\!\bS_1^c(\psi)+\delta_1S_0^c(\psi)+\cJ_1,\labell{S0122}\\
S_2^c(\psi)&=&\!\!\bS_2^c(\psi)+\delta_1S_1^c(\psi)-\frac{1}{2}\delta_1^2S_0^c(\psi)+\frac{1}{2}\delta_2S_0^c(\psi)+\cJ_2\,,\nn
\eeqa
where the one-dimensional field redefinitions $\psi \to \psi + \alpha' \delta \psi^{(1)} + \frac{1}{2}\alpha'^2 \delta \psi^{(2)} + \cdots$ on the $O(9,25)$-invariant action, and total derivative terms $\mathcal{J} = \sum_{m=0}^\infty \alpha'^m \mathcal{J}_m$ are used. In the above equations, $\delta_i$ on the action indicates that the Taylor expansion of the action contains the $i$-th order perturbation $\delta \psi^{(i)}$, and $\delta_1^2$ on the action means the Taylor expansion contains two first-order perturbations $\delta \psi^{(1)} \delta \psi^{(1)}$.
The one-dimensional actions on the left-hand side of the above equations in a particular scheme should be written in the canonical form given by \reef{cosmH}. In this paper, we are going to show that the couplings in the heterotic theory up to order $\alpha'^2$ that have been found by the truncated T-duality in \cite{Garousi:2024avb,Garousi:2024imy} can be written in the $O(9,25)$-invariant form given by \reef{cosmH} with the same coefficients that are fixed by the pure gravity couplings \cite{Codina:2021cxh}.

The paper is structured as follows:
In Section 2, we review the observation that the cosmological reduction of the leading order effective action of the heterotic string can be written in an $O(9,25)$-invariant form \cite{Maharana:1992my,Hohm:2014sxa}.
In Section 3, we show that the cosmological reduction of the four-derivative couplings that have been recently found in \cite{Garousi:2024avb} can be written as  $e^{-\phi}\tr(\dH^4)$ after using appropriate one-dimensional field redefinitions and total derivative terms. The coefficient of this term is the same as the one produced by studying pure gravity couplings.
In Section 4, we show that the cosmological reduction of the six-derivative couplings that have been found in \cite{Garousi:2024imy} and the six-derivative couplings that result from the Green-Schwarz mechanism \cite{Green:1984sg} (where $H \to H - (3\alpha'/2) \Omega$ is substituted into the 2- and 4-derivative couplings) becomes zero after using appropriate one-dimensional field redefinitions and total derivative terms. This is again the same result that one finds by studying the cosmological reduction of the pure gravity part.
Section 6 provides a concise discussion of our findings and their implications.
Throughout our calculations, we utilize the "xAct" package  \cite{Nutma:2013zea}  for computational purposes.

\section{Cosmological symmetry at the leading order}

In this section, we review the cosmological symmetry of the leading-order effective action of heterotic string theory.

The bosonic part of the leading order action is given as \cite{Gross:1985rr,Gross:1986mw}.
\beqa
{\bf S}_{0}&=&-\frac{2}{\kappa^2}\int d^{10}x \sqrt{-G}e^{-2\Phi}\Big[R-\frac{1}{12}H_{\alpha\beta\gamma}H^{\alpha\beta\gamma}+4\nabla_\alpha\Phi\nabla^\alpha\Phi-\frac{1}{4}F_{\mu\nu I}F^{\mu\nu I}\Big]\,,\labell{two2}
\eeqa
where the YM gauge field matrix is defined as ${\cal A}_{\mu} = A_{\mu}{}^I \lambda^I$, where the antisymmetric matrices $\lambda^I$ represent the adjoint representation of the gauge group $SO(32)$ or $E_8\times E_8$ with the normalization $\Tr(\lambda^I\lambda^J)=\delta^{IJ}$. The $H$-field strength without its Lorentz Chern-Simons contribution is \cite{Gross:1985rr,Gross:1986mw}:
\beqa
H_{\mu\nu\rho} &=& 3\partial_{[\mu}B_{\nu\rho]} - \frac{3}{2}A_{[\mu}{}^{I}F_{\nu\rho]}{}_{I}\,. \labell{H}
\eeqa
The $F$-field strength when the YM field belongs to the Cartan subalgebra where $[\lambda^I,\lambda^J]=0$ is given as
\beqa
F_{\mu\nu}{}^{I} &=& \partial_\mu A_{\nu}{}^{I} - \partial_\nu A_{\mu}{}^{I}. \labell{F}
\eeqa
We use cosmological reduction when the YM field is restricted to the Cartan subalgebra.

When the fields depend only on time, it is possible to write the metric, $B$-field, YM gauge field, and dilaton as follows:
 \beqa
G_{\mu\nu}=\left(\matrix{-n^2& 0&\cr 0&G_{ij}\!\!\!\!\!\!\!&}\right), B_{\mu\nu}= \left(\matrix{0&0\cr0&B_{ij}\!\!\!\!\!\!\!\!&}\right),   A_{\mu}{}^I= \left(\matrix{0&\cr A_i{}^I\!\!\!\!\!\!&}\right), 2\Phi=\phi+\frac{1}{2}\log\det(G_{ij})\,,\labell{creduce}\eeqa
where the lapse function $n(t)$ can also be fixed to $n=1$. The cosmological reduction of the action in \reef{two2} then becomes:
\beqa
\bS_0^c&=&-\frac{2}{\kappa^2}\int dt e^{-\phi}\Bigg[\frac{1}{2}\dA_i{}^I\dA_{iI}+\frac{1}{4}\dcB_{ij}\dcB^{ij}-\frac{3}{4}\dG_{ij}\dG^{ij}-G^{ij}\dG_{ij}\dP-\dP^2+G^{ij}\ddot{G}_{ij}\Bigg]\,,
\eeqa
where $\dG^{ij}\equiv G^{ik}G^{il}\dG_{kl}$, and $\dcB_{ij}$ is defined as 
\beqa
\dcB_{ij}\equiv\dB_{ij}+\frac{1}{2}A_i{}^I\dA_{jI}-\frac{1}{2}A_j{}^I\dA_{iI}\,.\labell{cB}
\eeqa
 Removing a  total derivative term,
one can write $\!\!\bS_0^c$ as
\beqa
S_0^c&=&\bS_0^c+\cJ_0\,=\,-\frac{2}{\kappa^2}\int dt e^{-\phi}\Bigg[\frac{1}{2}\dA_i{}^I\dA_{iI}+\frac{1}{4}\dcB_{ij}\dcB^{ij}+\frac{1}{4}\dG_{ij}\dG^{ij}-\dP^2\Bigg]\,.\labell{S0c}
\eeqa
The above one-dimensional action should be invariant under the $O(9,25)$-transformations. The dilaton $\phi$ is invariant, and the other functions should be combined into an $O(9,25)$-invariant expression.

Using the $O(9,25)$-valued generalized metric $\cH$ which is defined as $\cH=\eta \hat{\cH}$, where the $O(9,25)$-valued matrix $\hat{\cH}$ is given as \cite{Maharana:1992my,Hohm:2014sxa}.
\beqa
 \hat{\cH}\!\!\equiv\!\! \left(\matrix{ \hat{\cH}^{ij}& \hat{\cH}^{i}{}_j& \hat{\cH}^i{}_J&\cr  \hat{\cH}_i{}^j& \hat{\cH}_{ij}& \hat{\cH}_{iJ}&\cr  \hat{\cH}_I{}^j& \hat{\cH}_{Ij}& \hat{\cH}_{IJ}\!\!\!\!&}\!\!\!\!\right)\!\!=\!\! \left(\matrix{G^{ij}&-G^{ik}C_{kj}& -G^{ik}A_{kI}&\cr -C_{ki}G^{jk}&G_{ij}+C_{ki}G^{kl}C_{lj}+A_i{}^IA_{jI}&C_{ki}G^{kl}A_{lJ}+A_{iJ}&\cr -A_{kI}G^{jk}&C_{kj}G^{kl}A_{lI}+A_{jI}&\delta_{IJ}+A_{kI}G^{kl}A_{lJ}\!\!\!\!\!&}\!\!\!\right)\labell{S}
\eeqa
in which 
\beqa
C_{ij}&=&B_{ij}+\frac{1}{2}A_i{}^IA_{jI}\,,
\eeqa
and $\eta$ is the metric of the $O(9,25)$ group, which is
\beqa
\eta&=& \left(\matrix{0& \delta^i{}_j&0&\cr \delta_i{}^j&0&0&\cr 0&0&\delta_{IJ}&}\!\!\!\!\!\right).
\eeqa
One can calculate the $O(9,25)$-invariant expression involving the trace of the first time-derivative of $\cH$. One finds that the trace of an odd number of $\dH$ is zero, whereas the trace of an even number of $\dH$ is non-zero. The trace of two $\dH$ is
\beqa
\tr(\dH^2)&=&-4\dA_i{}^I\dA_{iI}-2\dcB_{ij}\dcB^{ij}-2\dG_{ij}\dG^{ij}\,.
\eeqa 
Then one can write the action \reef{S0c} in the following manifestly $O(9,25)$-invariant form:
\beqa
S_0^c&=&-\frac{2}{\kappa^2}\int dt e^{-\phi}\Bigg[-\dP^2-\frac{1}{8}\tr(\dH^2)\Bigg]\,.\labell{S0}
\eeqa
The $O(9,25)$-symmetry dictates that the lapse function can be included into the action \reef{S0} or \reef{S0c} by replacing $dt$ with $dt/n$  \cite{Hohm:2015doa}. This is needed for writing the cosmological reduction of higher derivative couplings in the canonical form of \reef{cosmH} because one needs to impose one-dimensional field redefinition of all functions, including the lapse function. Note that when the YM field is zero, the generalized metric $\cH$ reduces to the generalized metric $\cS$ considered in  \cite{Hohm:2015doa,Hohm:2019jgu}.

\section{Cosmological symmetry at order $\alpha'$}

In this section, we study the cosmological symmetry of the effective action of heterotic string theory at order $\alpha'$.

The four-derivative couplings and higher appear in many different forms that are connected through field redefinitions \cite{Metsaev:1987zx}. Using the truncated T-duality transformation on the four-derivative couplings in the maximal basis, in which the redundant terms due to field redefinitions are not removed, it has been found in \cite{Garousi:2024avb} that the effective action has 17 arbitrary parameters. Different values for these parameters correspond to different schemes for the effective action. One particular scheme for the coupling is \cite{Garousi:2024avb}:
\beqa
{\bf S}_1&=&-\frac{2}{8\kappa^2}\int d^{10}x \sqrt{-G}e^{-2\Phi}\Big[\frac{1}{4} F_{\alpha }{}^{\gamma J} F^{\alpha \beta I} 
F_{\beta }{}^{\delta }{}_{J} F_{\gamma \delta I} + 
\frac{1}{2} F_{\alpha }{}^{\gamma }{}_{I} F^{\alpha \beta 
I} F_{\beta }{}^{\delta J} F_{\gamma \delta J}\nn\\&& -  
\frac{1}{8} F_{\alpha \beta }{}^{J} F^{\alpha \beta I} F_{
\gamma \delta J} F^{\gamma \delta }{}_{I} -  \frac{1}{16} 
F_{\alpha \beta I} F^{\alpha \beta I} F_{\gamma \delta J} 
F^{\gamma \delta J} + \frac{1}{4} F^{\alpha \beta I} 
F^{\gamma \delta }{}_{I} H_{\alpha \gamma }{}^{\epsilon } 
H_{\beta \delta \epsilon } \nn\\&&-  \frac{1}{8} F^{\alpha \beta I} 
F^{\gamma \delta }{}_{I} H_{\alpha \beta }{}^{\epsilon } 
H_{\gamma \delta \epsilon } + \frac{1}{24} H_{\alpha 
}{}^{\delta \epsilon } H^{\alpha \beta \gamma } H_{\beta 
\delta }{}^{\varepsilon } H_{\gamma \epsilon \varepsilon } -  
\frac{1}{8} H_{\alpha \beta }{}^{\delta } H^{\alpha \beta 
\gamma } H_{\gamma }{}^{\epsilon \varepsilon } H_{\delta 
\epsilon \varepsilon } \nn\\&&+ \frac{1}{144} H_{\alpha \beta \gamma 
} H^{\alpha \beta \gamma } H_{\delta \epsilon \varepsilon } H^{
\delta \epsilon \varepsilon } + H_{\alpha }{}^{\gamma \delta } 
H_{\beta \gamma \delta } R^{\alpha \beta } - 4 
R_{\alpha \beta } R^{\alpha \beta } -  
\frac{1}{6} H_{\alpha \beta \gamma } H^{\alpha \beta \gamma } 
R + R^2 \nn\\&&+ R_{\alpha \beta \gamma 
\delta } R^{\alpha \beta \gamma \delta } -  
\frac{1}{2} H_{\alpha }{}^{\delta \epsilon } H^{\alpha \beta 
\gamma } R_{\beta \gamma \delta \epsilon } -  
\frac{2}{3} H_{\beta \gamma \delta } H^{\beta \gamma \delta } 
\nabla_{\alpha }\nabla^{\alpha }\Phi + \frac{2}{3} H_{\beta 
\gamma \delta } H^{\beta \gamma \delta } \nabla_{\alpha }\Phi 
\nabla^{\alpha }\Phi \nn\\&&+ 8 R \nabla_{\alpha }\Phi 
\nabla^{\alpha }\Phi - 16 R_{\alpha \beta } 
\nabla^{\alpha }\Phi \nabla^{\beta }\Phi + 16 \nabla_{\alpha 
}\Phi \nabla^{\alpha }\Phi \nabla_{\beta }\Phi \nabla^{\beta 
}\Phi - 32 \nabla^{\alpha }\Phi \nabla_{\beta }\nabla_{\alpha 
}\Phi \nabla^{\beta }\Phi \nn\\&&+ 2 H_{\alpha }{}^{\gamma \delta } 
H_{\beta \gamma \delta } \nabla^{\beta }\nabla^{\alpha }\Phi +2H^{\alpha\beta\gamma}\Omega_{\alpha\beta\gamma}\Big]\,.\labell{fourmax}
\eeqa
The Chern-Simons three-form  in the above action which is resulted from replacing $H\rightarrow H-(3\alpha'/2)\Omega$ into the 2-derivative action \reef{two2}, is given by:
\beqa
\Omega_{\mu\nu\alpha}&=&\omega_{[\mu {\mu_1}}{}^{\nu_1}\partial_\nu\omega_{\alpha] {\nu_1}}{}^{\mu_1}+\frac{2}{3}\omega_{[\mu {\mu_1}}{}^{\nu_1}\omega_{\nu {\nu_1}}{}^{\alpha_1}\omega_{\alpha]{\alpha_1}}{}^{\mu_1}\,\,;\,\,\,\omega_{\mu {\mu_1}}{}^{\nu_1}=e^\nu{}_{\mu_1}\nabla_\mu e_\nu{}^{\nu_1} \,,
\eeqa
where $e_\mu{}^{\mu_1}e_\nu{}^{\nu_1}\eta_{\mu_1\nu_1}=G_{\mu\nu}$. The covariant derivative in the definition of the spin connection applies only on the curved indices of the frame $e_\mu{}^{\mu_1}$. Our index convention is that $\mu, \nu, \ldots$ are the indices of the curved spacetime, and $\mu_1, \nu_1, \ldots$ are the indices of the flat tangent space.

For zero YM field, the above couplings become those that have been found by Meissner \cite{Meissner:1996sa} up to a total derivative term. In the above scheme, the propagators of the leading order action do not receive four-derivative corrections. It has been shown in \cite{Garousi:2024avb} that the above couplings are consistent with the sphere-level S-matrix elements.

Using the cosmological reductions \reef{creduce}, one can calculate the following cosmological reduction for different tensors that appear in the effective action in the above scheme and also in the arbitrary schemes found in \cite{Garousi:2024avb}:
\beqa
&&R_{ijkl}=-\frac{1}{4}\dG_{il}\dG_{jk}+\frac{1}{4}\dG_{ik}\dG_{jl};\,R_{i0jk}=0;\,R_{i0j0}=\frac{1}{4}\dG_{ik}\dG^k{}_j-\frac{1}{2}\ddot{G}_{ij};\,H_{ijk}=0\;\,H_{ij0}=\dcB_{ij};\nn\\
&&\nabla_l H_{ijk}=-\frac{1}{2}\dcB_{jk}\dG_{il}+\frac{1}{2}\dcB_{ik}\dG_{jl}-\frac{1}{2}\dcB_{ij}\dG_{kl};\,\nabla_0H_{ij0}=\frac{1}{2}\dcB_j{}^k\dG_{ik}-\frac{1}{2}\dcB_i{}^k\dG_{jk}+\ddot{\cB}_{ij};\nn\\
&&\nabla_0H_{ijk}=0\,\,;\,\, \nabla_k H_{ij0}=0;\,\nabla_0\phi=\dot{\phi};\,\nabla_i\phi=0;\,\nabla_0\nabla_i\phi=0;\,\nabla_0\nabla_0\phi=\ddot{\phi}\nn\\
&&\nabla_i\nabla_j\phi=-\frac{1}{2}\dot{\phi}\dG_{ij};\,\nabla_0G_{ij}=\dG_{ij};\,\nabla_iG_{jk}=0;\,\nabla_0\nabla_iG_{jk}=0;\,\nabla_0\nabla_0G_{ij}=\ddot{G}_{ij}\nn\\
&&\nabla_i\nabla_j G_{kl}=-\frac{1}{2}\dG_{kl}\dG_{ij};\,\Omega_{ijk}=0;\,\Omega_{ij0}=\frac{1}{12}\dG_j{}^k\ddot{G}_{ik}-\frac{1}{12}\dG_i{}^k\ddot{G}_{jk};\,F_{i0I}=-\dA_{iI};\, F_{ijI}=0\nn\\
&&\nabla_jF_{i0I}=0;\,\nabla_0F_{i0I}=\frac{1}{2}\dA^j{}_I\dG_{ij}-\ddot{A}_{iI};\,\nabla_k F_{ijI}=-\frac{1}{2}\dA_{jI}\dG_{ik}+\frac{1}{2}\dA_{iI}\dG_{jk};\,\nabla_0F_{ijI}=0.\labell{cRF}
\eeqa
Note that when the dilaton in the effective action is replaced by $\Phi=\frac{1}{2}\phi+\frac{1}{4}\log\det(G_{ij})$, then the spacetime derivatives of the scalars $\phi$ and $G_{ij}$ appear in the reduction. Hence, we have included these reductions into the above equation. 
In finding the above reduction for $\Omega$, one may assume the metric is diagonal, which is possible for cosmological reduction\footnote{Note that this assumption cannot be made for the circular reduction because the $U(1)$ gauge field appears in the off-diagonal of the metric.
}. Then one can relate the partial derivative of the frame to the partial derivative of the metric as $\partial_\alpha e_{\beta}{}^{\beta_1}=\frac{1}{2}\partial_\alpha G_{\beta\gamma} e^{\gamma\beta_1}$. In this case, $\Omega$ can be written in terms of only partial derivatives of the metric. For the cosmological case, in which the derivative indices must contract with themselves, one finds
\beqa
\Omega_{\alpha\beta\gamma}&=&-\frac{1}{2}G^{\delta\lambda}G^{\mu\nu}\prt_{[\beta}\prt_\delta G_{|\gamma|\nu}\prt_\lambda G_{\alpha]\mu}\,,
\eeqa
where the antisymmetrization of the indices is with respect to $\alpha,\beta,\gamma$. We have used this expression for calculating the cosmological reductions of the Chern-Simons three-form in \reef{cRF}.

Using the reduction \reef{cRF}, one finds the cosmological reduction of the action \reef{fourmax} to be
\beqa
\bS_1^c&=&-\frac{2}{8\kappa^2}\int dt e^{-\phi}\Bigg[\frac{1}{2} \dA_{i}{}^{J} \dA^{iI} \dA_{jJ} \dA^{j}{}_{I} + 
\frac{1}{4} \dA_{iI} \dA^{iI} \dA_{jJ} \dA^{jJ} -  \frac{3}{8} 
\dcB_{i}{}^{k} \dcB^{ij} \dcB_{j}{}^{l} \dcB_{kl} -  \frac{1}{16} 
\dcB_{ij} \dcB^{ij} \dcB_{kl} \dcB^{kl} \nn\\&&+ \frac{1}{4} \dcB^{ij} 
\dcB^{kl} \dG_{ik} \dG_{jl} + \frac{1}{2} \dcB_{i}{}^{k} \dcB^{ij} 
\dG_{j}{}^{l} \dG_{kl} -  \frac{7}{8} \dG_{i}{}^{k} \dG^{ij} 
\dG_{j}{}^{l} \dG_{kl} + \dG^{i}{}_{i} \dG_{j}{}^{l} \dG^{jk} 
\dG_{kl} -  \frac{1}{8} \dcB_{ij} \dcB^{ij} \dG_{kl} \dG^{kl} \nn\\&&+ 
\frac{7}{16} \dG_{ij} \dG^{ij} \dG_{kl} \dG^{kl} + \dcB_{i}{}^{k} 
\dcB^{ij} \dG_{jk} \dot{\phi} + \frac{5}{2} \dG^{i}{}_{i} \dG_{jk} 
\dG^{jk} \dot{\phi} + \frac{1}{2} \dcB_{ij} \dcB^{ij} \dot{\phi}^2 + 
\frac{5}{2} \dG_{ij} \dG^{ij} \dot{\phi}^2 + \dG^{i}{}_{i} 
\dG^{j}{}_{j} \dot{\phi}^2 \nn\\&&+ 2 \dG^{i}{}_{i} \dot{\phi}^3 + \dot{\phi}^4 - 2 
\dot{\phi}^2 \ddot{G}^{i}{}_{i} + \dcB^{ij} \dG_{i}{}^{k} \ddot{G}_{jk} + 
\dG_{i}{}^{k} \dG^{ij} \ddot{G}_{jk} -  \dG^{i}{}_{i} \dG^{jk} \ddot{G}_{jk} - 
2 \dG^{i}{}_{i} \dot{\phi} \ddot{G}^{j}{}_{j}\nn\\&& -  \frac{1}{2} \dG_{ij} 
\dG^{ij} \ddot{G}^{k}{}_{k} -  \dG^{i}{}_{i} \dG^{j}{}_{j} \ddot{\phi} - 4 
\dG^{i}{}_{i} \dot{\phi} \ddot{\phi} - 4 \dot{\phi}^2 \ddot{\phi} \Bigg].\labell{cS1}
\eeqa
Obviously, the above action is not invariant under $O(9,25)$ transformations. For example, it contains $\dG^i{}_i$, which is not invariant. According to the equation \reef{S0122}, to write it in $O(9,25)$-form, one must add total derivative terms to it and add the four-derivative terms resulting from using one-dimensional field redefinitions into the leading $O(9,25)$-invariant action \reef{S0c}.

We add the following total derivative terms:
\beqa
\cJ_1&=&-\frac{2}{\kappa^2}\int dt\frac{d}{dt}(e^{-\phi}\cI_1)\,,\labell{J1}
\eeqa
where  $\cI_1$ encompasses all possible terms at the three-derivative level, constructed from $\dA$, $\dot{\phi}$, $\dcB$, $\dG$, $\ddot{A}$, $\ddot{\phi}$, $\ddot{\cB}$, $\ddot{G}$, and so on, each with an arbitrary parameter.

To use field redefinition, one should perturb the fields in the $O(9,25)$-invariant action \reef{S0c}. We note that the expression $\dcB_{ij}$ which is defined in \reef{cB} is not the time-derivative of a field. It is in fact $H_{0ij}$ and $H_{\alpha\beta\gamma}$ is not an exact three-form. It satisfies instead the following Bianchi identity:
\beqa
\partial_{[\alpha}H_{\beta\mu\nu]}+\frac{3}{4}F_{[\alpha\beta}{}^{ij}F_{\mu\nu]}{}_{ij}&=&0\,.\labell{iden0}
\eeqa
This identity dictates that the perturbation of $H$ is related to the perturbation of the YM field $A$ and the perturbation of the $B$-field as \cite{Garousi:2024avb}.
\beqa
\delta H_{\mu\nu\alpha}=\prt_\mu\delta B_{\nu\alpha}+\prt_\alpha\delta B_{\mu\nu}+\prt_\nu\delta B_{\alpha\mu}-\delta A_\mu{}^I F_{\nu\alpha I}-\delta A_\alpha{}^IF_{\mu\nu I}-\delta A_\nu{}^IF_{\alpha\mu I}\,.
\eeqa
From this relation, one finds:
\beqa
\delta \dcB_{ij}&=&\frac{d}{dt}(\delta B_{ij})-(\dA_{iI}\delta A_j{}^I-\dA_{jI}\delta A_i{}^I)\,,\labell{tcB}
\eeqa
which relates the perturbation of $\dcB_{ij}$ to the perturbations of $B$ and $A$.

Now, if we perturb the variables in  \reef{creduce}, including the lapse function, as:
\begin{eqnarray}
G_{ij} &\rightarrow &G_{ij}+\alpha' \delta G^{(1)}_{ij},\nn\\
B_{ij} &\rightarrow &B_{ij} + \alpha'\delta B^{(1)}_{ij},\nn\\
A_{iI} &\rightarrow &A_{iI} + \alpha'\delta A^{(1)}_{iI},\nn\\
\phi &\rightarrow &\phi + \alpha'\delta\phi^{(1)},\nn\\
n &\rightarrow &n + \alpha' \delta n^{(1)},\labell{gbpn}
\end{eqnarray}
where the matrices $\delta G^{(1)}_{ij}$, $\delta B^{(1)}_{ij}$, $\delta A^{(1)}_{iI}$, $\delta\phi^{(1)}$, and $\delta n^{(1)}$ consist of all possible terms at the 2-derivative level, constructed from $\dP$, $\dB$, $\dG$, $\dA$, $\ddot{\phi}$, $\ddot{B}$, $\ddot{G}$, $\ddot{A}$  with arbitrary coefficients, then the Taylor expansion of the action \reef{S0c} produces the following four-derivative terms:
\beqa
\delta_1 S_0^c&=&-\frac{2}{\kappa^2}\int dt e^{-\phi}\Bigg[\delta n^{(1)}\left(-\frac{1}{2}\dA_{iI}\dA^{iI}-\frac{1}{4}\dB_{ij}\dB^{ij}-\frac{1}{4}\dG_{ij}\dG^{ij}+\dP^2\right)\nn\\
&&+\delta \phi^{(1)}\left(-\frac{1}{2}\dA_{iI}\dA^{iI}-\frac{1}{4}\dB_{ij}\dB^{ij}-\frac{1}{4}\dG_{ij}\dG^{ij}+\dP^2\right)-2\dP\frac{d}{dt}\delta \phi^{(1)}\nn\\
&&+\delta G^{(1)}_{ij}\left(-\frac{1}{2}\dA^{iI}\dA^j{}_{I}-\frac{1}{2}\dB_k{}^j\dB^{ki}-\frac{1}{2}\dG_k{}^j\dG^{ki}\right)+\frac{1}{2}\dG^{ij}\frac{d}{dt}\delta G^{(1)}_{ij}\nn\\
&&+\frac{1}{2}\dcB^{ij}\frac{d}{dt}\delta B^{(1)}_{ij}-\dA^{iI}\dcB_i{}^j\delta A^{(1)}_{jI}+\dA^{iI}\frac{d}{dt} \delta A^{(1)}_{iI}\Bigg]\,,\labell{dS0c}
\eeqa
as well as six-derivative terms that we consider in the next section when studying six-derivative coupling. It also produces eight-derivative terms and higher, which we are not interested in for this paper.

Adding the total derivative terms \reef{J1} and the above field redefinition terms to the one-dimensional action \reef{cS1}, one can express $\bS_1^c+\delta_1 S_0^c+\cJ_1$ in various $O(9,25)$-invariant forms. However, if one wishes to write it in the canonical form given by \reef{cosmH}, certain constraints need to be imposed. These constraints require that $\bS_1^c+\delta_1 S_0^c+\cJ_1$  only includes first derivative terms, does not contain derivatives of the dilaton, and does not include the trace of two $\dB$, $\dA$ or $\dG$ terms.
By imposing these constraints, one finds that the following field redefinitions can be employed:
 \beqa
\delta n^{(1)}&=&\frac{1}{8}\Big(\frac{1}{2}\dA_{iI}\dA^{iI}- \frac{1}{4} \dB_{ij} \dB^{ij} - \frac{1}{4} \dG_{ij} \dG^{ij}+ \frac{1}{2}\dP^2\Big),\nn\\
\delta \phi^{(1)}&=&\frac{1}{16} \dP^2,\nn\\
\delta A^{(1)}_{iI}&=&0,\nn\\
\delta G^{(1)}_{ij}&=&-\frac{1}{8}( \dG_i{}^k\dG_{jk}+ \dcB_i{}^k\dcB_{jk}+ \dcB_i{}^k\dG_{jk} +\dG_i{}^k\dcB_{jk}),\nn\\
\delta B^{(1)}_{ij}&=&\frac{1}{8}\Big(\dcB_{j}{}^{k} \dG_{ik} -  \dcB_{i}{}^{k} \dG_{jk} \Big) .\labell{dG1dB1}
\eeqa
For zero YM field, the above transformations are those that have been found in \cite{Gholian:2023kjj}\footnote{Note that the sign of the $\Omega$-term in \reef{fourmax} is different from the corresponding term in \cite{Gholian:2023kjj}. This is why the sign of the last two terms in $\delta G^{(1)}_{ij}$ in equation \reef{dG1dB1} is different from the corresponding terms in \cite{Gholian:2023kjj}.}.
Using the above field redefinitions, one finds that, up to some total derivative terms, the cosmological action at order $\alpha'$ can be expressed as follows:
 \beqa
 S_1^c&=&\!\!\bS_1^c+\delta_1 S_0^c+\cJ_1\,=\,-\frac{2}{8\kappa^2}\int dt e^{-\phi}\Bigg[ \frac{1}{2} \dA_{i}{}^{J} \dA^{iI} \dA_{jJ} \dA^{j}{}_{I} + \frac{1}{2} \dA^{iI} 
\dA^{j}{}_{I} \dcB_{i}{}^{k} \dcB_{jk}  \nn\\&&+ 
 \dA^{iI} \dA^{j}{}_{I} \dcB_{i}{}^{k} 
\dG_{jk} + \frac{1}{2} \dA^{iI} \dA^{j}{}_{I} \dG_{i}{}^{k} \dG_{jk}+\frac{1}{8} \dcB_{i}{}^{k} \dcB^{ij} \dcB_{j}{}^{l} \dcB_{kl} \nn\\&&-  
\frac{1}{4} \dcB^{ij} \dcB^{kl} \dG_{ik} \dG_{jl}+ \frac{1}{2} 
\dcB_{i}{}^{k} \dcB^{ij} \dG_{j}{}^{l} \dG_{kl} + \frac{1}{8} 
\dG_{i}{}^{k} \dG^{ij} \dG_{j}{}^{l} \dG_{kl}\Bigg]\,.\labell{action2}
 \eeqa
The total derivative terms can be ignored as they have no impact on the calculations at order $\alpha'^2$.

Now, by utilizing the definition of the generalized metric in  \reef{S}, we can find the following $O(9,25)$-invariant expression:
\beqa
\tr(\dH^4)&\!\!\!\!=\!\!\!\!\! & 8 \dA_{i}{}^{J} \dA^{iI} \dA_{jJ} \dA^{j}{}_{I} + 8 \dA^{iI} 
\dA^{j}{}_{I} \dcB_{i}{}^{k} \dcB_{jk} + 2 \dcB_{i}{}^{k} \dcB^{ij} 
\dcB_{j}{}^{l} \dcB_{kl} + 16 \dA^{iI} \dA^{j}{}_{I} \dcB_{i}{}^{k} 
\dG_{jk} + 8 \dA^{iI} \dA^{j}{}_{I} \dG_{i}{}^{k} \dG_{jk}\nn\\&\!\!\!\!\!\!& - 4 
\dcB^{ij} \dcB^{kl} \dG_{ik} \dG_{jl} + 8 \dcB_{i}{}^{k} \dcB^{ij} 
\dG_{j}{}^{l} \dG_{kl} + 2 \dG_{i}{}^{k} \dG^{ij} \dG_{j}{}^{l} 
\dG_{kl}\,.
\eeqa
 Using this $O(9,25)$-invariant expressions, we can express \reef{action2} as: 
\beqa
 S_1^c&=&-\frac{2}{2^7\kappa^2}\int dt e^{-\phi}\tr(\dH^4)\,.\labell{action3}
 \eeqa
These results coincide with those found in \cite{Codina:2021cxh} through the study of the pure gravity parts of the couplings. The lapse function can be incorporated into the $O(9,25)$-invariant action \reef{action3} or  \reef{action2} by replacing $dt$ with $dt/n^3$.

We have done the same calculations for the four-derivative couplings in an arbitrary scheme found in \cite{Garousi:2024avb} that incorporate 17 arbitrary parameters, and found exactly the same result as in \reef{action3}  but with one-dimensional field redefinitions which are different than those in \reef{dG1dB1}. This confirms that the four-derivative couplings found in \cite{Garousi:2024avb} satisfy the expected cosmological symmetry.

\section{Cosmological symmetry at order $\alpha'^2$}

In this section, we study the cosmological symmetry of the effective action of heterotic string theory at order $\alpha'^2$.

It has been shown in \cite{Garousi:2024imy} that using the truncated T-duality transformations on the six-derivative couplings in the maximal basis which has 801 couplings, and the six-derivative couplings resulting from the replacement of $H \rightarrow H - (3\alpha'/2) \Omega$ into the fixed 2- and 4-derivative actions, one finds that T-duality produces 468 relations between these couplings. It has also been observed that the remaining 333 parameters are not physical parameters because they can all be removed by field redefinitions. The couplings that are found by the T-duality have then been written in a canonical form, after using field redefinitions, integration by parts, and Bianchi identities. They are  the following 85 couplings \cite{Garousi:2024imy}:
\beqa
{\bf S_2}^{({1})}&\!\!\!\!=\!\!\!\!&-\frac{2}{8^2\kappa^2}\int d^{10}x \sqrt{-G}e^{-2\Phi}\Big[[{\rm NS}\!\!-\!\!{\rm NS}]_{10}+[F^4H^2]_{12}+[F^2H^4]_{6}+[F^2H^2R]_{9}+[F^3H\nabla F]_{15}\nn\\&&+[F^4 R]_{4}+[F^2H^2\nabla H]_{8}+[F^2R\nabla H ]_{3}+[F^2(\nabla F)^2]_{9}+[F^2(\nabla H)^2]_{5}+[F^2R^2]_{4}\Big]\,,\labell{L22}
\eeqa
where 
\beqa
[{\rm NS}\!\!-\!\!{\rm NS}]_{10}&\!\!\!=\!\!\!& 2 H^{\alpha 
\beta \gamma } H^{\delta \epsilon \varepsilon } 
R_{\alpha \beta \delta }{}^{\mu } R_{\gamma 
\mu \epsilon \varepsilon }- \frac{1}{12} H_{\alpha }{}^{\delta \epsilon } H^{\alpha 
\beta \gamma } H_{\beta \delta }{}^{\varepsilon } H_{\gamma 
}{}^{\mu \zeta } H_{\epsilon \mu }{}^{\eta } H_{\varepsilon 
\zeta \eta } \nn\\&\!\!\!\!\!\!\!\!&- 2 H_{\alpha }{}^{\delta \epsilon } H^{\alpha 
\beta \gamma } R_{\beta \delta }{}^{\varepsilon \mu } 
R_{\gamma \varepsilon \epsilon \mu }  - 2 H_{\alpha }{}^{\delta \epsilon } 
H^{\alpha \beta \gamma } R_{\beta }{}^{\varepsilon 
}{}_{\gamma }{}^{\mu } R_{\delta \varepsilon \epsilon 
\mu } \nn\\&\!\!\!\!\!\!\!\!&+ H_{\alpha }{}^{\delta \epsilon } H^{\alpha \beta 
\gamma } H_{\beta \delta }{}^{\varepsilon } H_{\gamma }{}^{\mu 
\zeta } R_{\epsilon \varepsilon \mu \zeta } - 4 
H^{\alpha \beta \gamma } H^{\delta \epsilon \varepsilon } 
R_{\gamma \epsilon \varepsilon \mu } \nabla_{\beta 
}H_{\alpha \delta }{}^{\mu }\nn\\&\!\!\!\!\!\!\!\!& -  H_{\alpha }{}^{\delta \epsilon 
} H^{\alpha \beta \gamma } R_{\delta \epsilon 
\varepsilon \mu } \nabla_{\gamma }H_{\beta }{}^{\varepsilon 
\mu } -  \frac{1}{2} H^{\alpha \beta \gamma } H^{\delta 
\epsilon \varepsilon } \nabla_{\beta }H_{\alpha \delta 
}{}^{\mu } \nabla_{\varepsilon }H_{\gamma \epsilon \mu }\nn\\&\!\!\!\!\!\!\!\!& -  
\frac{1}{2} H_{\alpha }{}^{\delta \epsilon } H^{\alpha \beta 
\gamma } H_{\beta \delta }{}^{\varepsilon } H_{\gamma }{}^{\mu 
\zeta } \nabla_{\varepsilon }H_{\epsilon \mu \zeta } + 
\frac{1}{4} H_{\alpha }{}^{\delta \epsilon } H^{\alpha \beta 
\gamma } \nabla_{\epsilon }H_{\delta \varepsilon \mu } 
\nabla^{\mu }H_{\beta \gamma }{}^{\varepsilon }\,,\nn\\
{[}F^4H^2{]}_{12}&=&\frac{1}{8} F_{\alpha }{}^{\gamma J} F^{\alpha \beta I} 
F^{\delta \epsilon }{}_{I} F^{\varepsilon \mu }{}_{J} 
H_{\beta \varepsilon \mu } H_{\gamma \delta \epsilon } + 
\frac{3}{2} F_{\alpha }{}^{\gamma J} F^{\alpha \beta I} 
F^{\delta \epsilon }{}_{I} F^{\varepsilon \mu }{}_{J} 
H_{\beta \delta \varepsilon } H_{\gamma \epsilon \mu }\nn\\&& -  
\frac{3}{2} F_{\alpha }{}^{\gamma }{}_{I} F^{\alpha \beta 
I} F^{\delta \epsilon J} F^{\varepsilon \mu }{}_{J} 
H_{\beta \delta \varepsilon } H_{\gamma \epsilon \mu } -  
\frac{1}{2} F_{\alpha }{}^{\gamma J} F^{\alpha \beta I} 
F^{\delta \epsilon }{}_{I} F^{\varepsilon \mu }{}_{J} 
H_{\beta \delta \epsilon } H_{\gamma \varepsilon \mu }\nn\\&& -  
\frac{1}{4} F_{\alpha }{}^{\gamma }{}_{I} F^{\alpha \beta 
I} F^{\delta \epsilon J} F^{\varepsilon \mu }{}_{J} 
H_{\beta \delta \epsilon } H_{\gamma \varepsilon \mu } -  
F_{\alpha }{}^{\gamma J} F^{\alpha \beta I} F_{\delta }{}^{
\varepsilon }{}_{J} F^{\delta \epsilon }{}_{I} H_{\beta 
\epsilon }{}^{\mu } H_{\gamma \varepsilon \mu }\nn\\&& - 2 F_{\alpha 
}{}^{\gamma }{}_{I} F^{\alpha \beta I} F_{\delta 
}{}^{\varepsilon }{}_{J} F^{\delta \epsilon J} H_{\beta 
\epsilon }{}^{\mu } H_{\gamma \varepsilon \mu } - 2 F_{\alpha 
}{}^{\gamma }{}_{I} F^{\alpha \beta I} F_{\beta }{}^{\delta 
J} F^{\epsilon \varepsilon }{}_{J} H_{\gamma \epsilon 
}{}^{\mu } H_{\delta \varepsilon \mu }\nn\\&& + \frac{1}{8} 
F_{\alpha \beta I} F^{\alpha \beta I} F^{\gamma \delta J} 
F^{\epsilon \varepsilon }{}_{J} H_{\gamma \epsilon }{}^{\mu } 
H_{\delta \varepsilon \mu } -  \frac{1}{16} F_{\alpha \beta 
I} F^{\alpha \beta I} F^{\gamma \delta J} F^{\epsilon 
\varepsilon }{}_{J} H_{\gamma \delta }{}^{\mu } H_{\epsilon 
\varepsilon \mu } \nn\\&&+ \frac{1}{2} F_{\alpha }{}^{\gamma }{}_{I} 
F^{\alpha \beta I} F_{\beta }{}^{\delta J} F_{\gamma 
}{}^{\epsilon }{}_{J} H_{\delta }{}^{\varepsilon \mu } 
H_{\epsilon \varepsilon \mu } + \frac{1}{8} F_{\alpha \beta 
I} F^{\alpha \beta I} F_{\gamma }{}^{\epsilon }{}_{J} 
F^{\gamma \delta J} H_{\delta }{}^{\varepsilon \mu } 
H_{\epsilon \varepsilon \mu }\,,\nn\\
{[}F^2H^4{]}_{6} &=&- F^{\alpha \beta I} F^{\gamma \delta }{}_{I} H_{\alpha 
\gamma }{}^{\epsilon } H_{\beta }{}^{\varepsilon \mu } 
H_{\delta \varepsilon }{}^{\zeta } H_{\epsilon \mu \zeta } + 
\frac{1}{2} F^{\alpha \beta I} F^{\gamma \delta }{}_{I} H_{
\alpha \beta }{}^{\epsilon } H_{\gamma }{}^{\varepsilon \mu } 
H_{\delta \varepsilon }{}^{\zeta } H_{\epsilon \mu \zeta }\nn\\&& -  
\frac{1}{2} F_{\alpha }{}^{\gamma }{}_{I} F^{\alpha \beta 
I} H_{\beta }{}^{\delta \epsilon } H_{\gamma }{}^{\varepsilon 
\mu } H_{\delta \varepsilon }{}^{\zeta } H_{\epsilon \mu \zeta 
} + \frac{1}{48} F_{\alpha \beta I} F^{\alpha \beta I} 
H_{\gamma }{}^{\varepsilon \mu } H^{\gamma \delta \epsilon } 
H_{\delta \varepsilon }{}^{\zeta } H_{\epsilon \mu \zeta } \nn\\&&-  
\frac{1}{2} F_{\alpha }{}^{\gamma }{}_{I} F^{\alpha \beta 
I} H_{\beta }{}^{\delta \epsilon } H_{\gamma \delta 
}{}^{\varepsilon } H_{\epsilon }{}^{\mu \zeta } H_{\varepsilon 
\mu \zeta } + \frac{1}{16} F_{\alpha \beta I} F^{\alpha 
\beta I} H_{\gamma \delta }{}^{\varepsilon } H^{\gamma \delta 
\epsilon } H_{\epsilon }{}^{\mu \zeta } H_{\varepsilon \mu 
\zeta }\,,\nn\\
{[}F^2H^2R{]}_{9}&=&2 F^{\alpha \beta I} F^{\gamma \delta }{}_{I} H_{\alpha 
}{}^{\epsilon \varepsilon } H_{\gamma \epsilon }{}^{\mu } 
R_{\beta \delta \varepsilon \mu } + F^{\alpha \beta 
I} F^{\gamma \delta }{}_{I} H_{\alpha \gamma }{}^{\epsilon } 
H_{\epsilon }{}^{\varepsilon \mu } R_{\beta \delta 
\varepsilon \mu } \nn\\&& -  F^{\alpha \beta I} F^{\gamma \delta 
}{}_{I} H_{\alpha }{}^{\epsilon \varepsilon } H_{\beta \epsilon 
}{}^{\mu } R_{\gamma \delta \varepsilon \mu } -  
\frac{1}{2} F^{\alpha \beta I} F^{\gamma \delta }{}_{I} H_{
\alpha \beta }{}^{\epsilon } H_{\epsilon }{}^{\varepsilon \mu } 
R_{\gamma \delta \varepsilon \mu }  \nn\\&&- 6 F_{\alpha }{}^{
\gamma }{}_{I} F^{\alpha \beta I} H_{\beta }{}^{\delta 
\epsilon } H_{\delta }{}^{\varepsilon \mu } R_{\gamma 
\epsilon \varepsilon \mu } + 2 F^{\alpha \beta I} F^{\gamma 
\delta }{}_{I} H_{\alpha \gamma }{}^{\epsilon } H_{\beta 
}{}^{\varepsilon \mu } R_{\delta \epsilon \varepsilon 
\mu } \nn\\&& + 2 F_{\alpha }{}^{\gamma }{}_{I} F^{\alpha \beta I} 
H_{\beta }{}^{\delta \epsilon } H_{\gamma }{}^{\varepsilon \mu 
} R_{\delta \epsilon \varepsilon \mu } -  \frac{3}{4} 
F_{\alpha \beta I} F^{\alpha \beta I} H_{\gamma 
}{}^{\varepsilon \mu } H^{\gamma \delta \epsilon } 
R_{\delta \epsilon \varepsilon \mu }  \nn\\&&- 2 F^{\alpha 
\beta I} F^{\gamma \delta }{}_{I} H_{\alpha \beta 
}{}^{\epsilon } H_{\gamma }{}^{\varepsilon \mu } 
R_{\delta \varepsilon \epsilon \mu }\,,\nn\\
{[}F^4 R{]}_{4}&=&2 F_{\alpha }{}^{\gamma J} F^{\alpha \beta I} F_{\delta 
}{}^{\varepsilon }{}_{J} F^{\delta \epsilon }{}_{I} 
R_{\beta \epsilon \gamma \varepsilon } + 2 F_{\alpha 
}{}^{\gamma }{}_{I} F^{\alpha \beta I} F_{\delta 
}{}^{\varepsilon }{}_{J} F^{\delta \epsilon J} 
R_{\beta \epsilon \gamma \varepsilon }\nn\\&& -  \frac{1}{4} 
F_{\alpha \beta I} F^{\alpha \beta I} F^{\gamma \delta J} 
F^{\epsilon \varepsilon }{}_{J} R_{\gamma \delta 
\epsilon \varepsilon } - 2 F_{\alpha }{}^{\gamma J} F^{\alpha 
\beta I} F_{\beta }{}^{\delta }{}_{J} F^{\epsilon 
\varepsilon }{}_{I} R_{\gamma \epsilon \delta 
\varepsilon }\,,\nn\\
{[}F^2H^2\nabla H{]}_{8}&=&-2 F^{\alpha \beta I} F^{\gamma \delta }{}_{I} H_{\alpha 
}{}^{\epsilon \varepsilon } H_{\gamma \epsilon }{}^{\mu } 
\nabla_{\delta }H_{\beta \varepsilon \mu } -  \frac{3}{4} F^{
\alpha \beta I} F^{\gamma \delta }{}_{I} H_{\alpha \gamma 
}{}^{\epsilon } H_{\epsilon }{}^{\varepsilon \mu } 
\nabla_{\delta }H_{\beta \varepsilon \mu }  \nn\\&&+ F^{\alpha \beta 
I} F^{\gamma \delta }{}_{I} H_{\alpha }{}^{\epsilon 
\varepsilon } H_{\beta \epsilon }{}^{\mu } \nabla_{\delta 
}H_{\gamma \varepsilon \mu } + \frac{3}{8} F^{\alpha \beta 
I} F^{\gamma \delta }{}_{I} H_{\alpha \beta }{}^{\epsilon } 
H_{\epsilon }{}^{\varepsilon \mu } \nabla_{\delta }H_{\gamma 
\varepsilon \mu }  \nn\\&&+ \frac{1}{2} F^{\alpha \beta I} 
F^{\gamma \delta }{}_{I} H_{\alpha \gamma }{}^{\epsilon } 
H_{\beta }{}^{\varepsilon \mu } \nabla_{\delta }H_{\epsilon 
\varepsilon \mu } + 2 F_{\alpha }{}^{\gamma }{}_{I} F^{\alpha 
\beta I} H_{\beta }{}^{\delta \epsilon } H_{\delta 
}{}^{\varepsilon \mu } \nabla_{\epsilon }H_{\gamma \varepsilon 
\mu }  \nn\\&&-  \frac{1}{2} F^{\alpha \beta I} F^{\gamma \delta 
}{}_{I} H_{\alpha \gamma }{}^{\epsilon } H_{\beta 
}{}^{\varepsilon \mu } \nabla_{\epsilon }H_{\delta \varepsilon 
\mu } -  \frac{1}{2} F^{\alpha \beta I} F^{\gamma \delta 
}{}_{I} H_{\alpha \beta }{}^{\epsilon } H_{\gamma 
}{}^{\varepsilon \mu } \nabla_{\mu }H_{\delta \epsilon 
\varepsilon }\,,\nn\\
{[}F^3H\nabla F{]}_{15}&=&- F^{\alpha \beta I} F^{\gamma \delta }{}_{I} F^{\epsilon 
\varepsilon J} H_{\delta \epsilon \varepsilon } \nabla_{\beta 
}F_{\alpha \gamma J} -  F^{\alpha \beta I} F^{\gamma 
\delta }{}_{I} F^{\epsilon \varepsilon J} H_{\gamma \delta 
\varepsilon } \nabla_{\beta }F_{\alpha \epsilon J} \nn\\&&- 2 
F_{\alpha }{}^{\gamma J} F^{\alpha \beta I} F^{\delta 
\epsilon }{}_{I} H_{\gamma \epsilon \varepsilon } 
\nabla_{\beta }F_{\delta }{}^{\varepsilon }{}_{J} + 4 
F_{\alpha }{}^{\gamma }{}_{I} F^{\alpha \beta I} F^{\delta 
\epsilon J} H_{\gamma \epsilon \varepsilon } \nabla_{\beta 
}F_{\delta }{}^{\varepsilon }{}_{J} \nn\\&&+ 6 F^{\alpha \beta I} 
F^{\gamma \delta }{}_{I} F^{\epsilon \varepsilon J} H_{\beta 
\delta \varepsilon } \nabla_{\gamma }F_{\alpha \epsilon J} + 
2 F_{\alpha }{}^{\gamma }{}_{I} F^{\alpha \beta I} 
F^{\delta \epsilon J} H_{\delta \epsilon \varepsilon } 
\nabla_{\gamma }F_{\beta }{}^{\varepsilon }{}_{J} \nn\\&&- 2 
F_{\alpha }{}^{\gamma J} F^{\alpha \beta I} F^{\delta 
\epsilon }{}_{I} H_{\beta \epsilon \varepsilon } 
\nabla_{\gamma }F_{\delta }{}^{\varepsilon }{}_{J} -  
F_{\alpha }{}^{\gamma J} F^{\alpha \beta I} F_{\beta 
}{}^{\delta }{}_{J} H_{\delta \epsilon \varepsilon } 
\nabla_{\gamma }F^{\epsilon \varepsilon }{}_{I} \nn\\&&+ 3 F_{\alpha 
}{}^{\gamma }{}_{I} F^{\alpha \beta I} F_{\beta }{}^{\delta 
J} H_{\delta \epsilon \varepsilon } \nabla_{\gamma 
}F^{\epsilon \varepsilon }{}_{J} -  \frac{1}{4} F_{\alpha 
\beta I} F^{\alpha \beta I} F^{\gamma \delta J} H_{\delta 
\epsilon \varepsilon } \nabla_{\gamma }F^{\epsilon \varepsilon 
}{}_{J}\nn\\&& - 6 F_{\alpha }{}^{\gamma }{}_{I} F^{\alpha \beta I} 
F^{\delta \epsilon J} H_{\gamma \epsilon \varepsilon } 
\nabla_{\delta }F_{\beta }{}^{\varepsilon }{}_{J} + 2 
F_{\alpha }{}^{\gamma J} F^{\alpha \beta I} F^{\delta 
\epsilon }{}_{I} H_{\beta \epsilon \varepsilon } 
\nabla_{\delta }F_{\gamma }{}^{\varepsilon }{}_{J} \nn\\&&+ 
F_{\alpha }{}^{\gamma }{}_{I} F^{\alpha \beta I} F_{\beta 
}{}^{\delta J} H_{\gamma \epsilon \varepsilon } \nabla_{\delta 
}F^{\epsilon \varepsilon }{}_{J} - 2 F_{\alpha }{}^{\gamma J} 
F^{\alpha \beta I} F^{\delta \epsilon }{}_{I} H_{\beta 
\gamma \varepsilon } \nabla_{\epsilon }F_{\delta 
}{}^{\varepsilon }{}_{J} \nn\\&&- 4 F^{\alpha \beta I} F^{\gamma 
\delta }{}_{I} F^{\epsilon \varepsilon J} H_{\beta \gamma 
\delta } \nabla_{\varepsilon }F_{\alpha \epsilon J}\,,\nn\\
{[}F^2R\nabla H {]}_{3}&=&-2 F^{\alpha \beta I} F^{\gamma \delta }{}_{I} 
R_{\gamma \delta \epsilon \varepsilon } \nabla_{\beta 
}H_{\alpha }{}^{\epsilon \varepsilon } + 4 F^{\alpha \beta I} 
F^{\gamma \delta }{}_{I} R_{\beta \delta \epsilon 
\varepsilon } \nabla_{\gamma }H_{\alpha }{}^{\epsilon 
\varepsilon } \nn\\&&- 8 F_{\alpha }{}^{\gamma }{}_{I} F^{\alpha 
\beta I} R_{\gamma \delta \epsilon \varepsilon } 
\nabla^{\varepsilon }H_{\beta }{}^{\delta \epsilon }\,,\nn\\
{[}F^2(\nabla F)^2{]}_{9}&=&-2 F_{\alpha }{}^{\gamma }{}_{I} F^{\alpha \beta I} 
\nabla_{\beta }F^{\delta \epsilon J} \nabla_{\gamma 
}F_{\delta \epsilon J} + \frac{1}{2} F^{\alpha \beta I} 
F^{\gamma \delta }{}_{I} \nabla_{\epsilon }F_{\gamma \delta 
J} \nabla^{\epsilon }F_{\alpha \beta }{}^{J} \nn\\&&+ 4 F^{\alpha 
\beta I} F^{\gamma \delta J} \nabla_{\beta }F_{\delta 
\epsilon J} \nabla^{\epsilon }F_{\alpha \gamma I} + 4 
F^{\alpha \beta I} F^{\gamma \delta J} \nabla_{\epsilon 
}F_{\beta \delta J} \nabla^{\epsilon }F_{\alpha \gamma I}\nn\\&& - 
 F^{\alpha \beta I} F^{\gamma \delta }{}_{I} 
\nabla_{\epsilon }F_{\beta \delta J} \nabla^{\epsilon 
}F_{\alpha \gamma }{}^{J} - 4 F_{\alpha }{}^{\gamma J} 
F^{\alpha \beta I} \nabla_{\delta }F_{\gamma \epsilon J} 
\nabla^{\epsilon }F_{\beta }{}^{\delta }{}_{I} \nn\\&&+ 4 F_{\alpha 
}{}^{\gamma J} F^{\alpha \beta I} \nabla_{\epsilon 
}F_{\gamma \delta J} \nabla^{\epsilon }F_{\beta }{}^{\delta 
}{}_{I} + 2 F_{\alpha }{}^{\gamma }{}_{I} F^{\alpha \beta I} 
\nabla_{\epsilon }F_{\gamma \delta J} \nabla^{\epsilon 
}F_{\beta }{}^{\delta J}\nn\\&& + \frac{1}{4} F_{\alpha \beta I} 
F^{\alpha \beta I} \nabla_{\epsilon }F_{\gamma \delta J} 
\nabla^{\epsilon }F^{\gamma \delta J}\,,\nn\\
{[}F^2(\nabla H)^2{]}_{5}&=&- \frac{2}{3} F_{\alpha }{}^{\gamma }{}_{I} F^{\alpha \beta 
I} \nabla_{\beta }H^{\delta \epsilon \varepsilon } 
\nabla_{\gamma }H_{\delta \epsilon \varepsilon } + F^{\alpha 
\beta I} F^{\gamma \delta }{}_{I} \nabla_{\beta }H_{\alpha 
}{}^{\epsilon \varepsilon } \nabla_{\delta }H_{\gamma \epsilon 
\varepsilon } \nn\\&&+ 2 F^{\alpha \beta I} F^{\gamma \delta 
}{}_{I} \nabla_{\delta }H_{\beta \epsilon \varepsilon } 
\nabla^{\varepsilon }H_{\alpha \gamma }{}^{\epsilon } - 2 
F_{\alpha }{}^{\gamma }{}_{I} F^{\alpha \beta I} 
\nabla_{\epsilon }H_{\gamma \delta \varepsilon } 
\nabla^{\varepsilon }H_{\beta }{}^{\delta \epsilon } \nn\\&&+ 
\frac{1}{2} F_{\alpha \beta I} F^{\alpha \beta I} \nabla_{
\epsilon }H_{\gamma \delta \varepsilon } \nabla^{\varepsilon 
}H^{\gamma \delta \epsilon }\,,\nn\\
{[}F^2R^2{]}_{4}&=&-2 F^{\alpha \beta I} F^{\gamma \delta }{}_{I} 
R_{\alpha \gamma }{}^{\epsilon \varepsilon } 
R_{\beta \delta \epsilon \varepsilon } + F^{\alpha 
\beta I} F^{\gamma \delta }{}_{I} R_{\alpha \beta 
}{}^{\epsilon \varepsilon } R_{\gamma \delta \epsilon 
\varepsilon } \nn\\&&- 4 F_{\alpha }{}^{\gamma }{}_{I} F^{\alpha 
\beta I} R_{\beta }{}^{\delta \epsilon \varepsilon } 
R_{\gamma \delta \epsilon \varepsilon } + \frac{1}{2} 
F_{\alpha \beta I} F^{\alpha \beta I} R_{\gamma 
\delta \epsilon \varepsilon } R^{\gamma \delta 
\epsilon \varepsilon }\,.
\eeqa
The 6-derivative couplings that result from replacing $H \rightarrow H - (3\alpha'/2) \Omega$ into the 2-derivative action \reef{two2} are:
\beqa
{\bf S_2}^{({2})}&=&-\frac{2}{\kappa^2}\int d^{10} x \sqrt{-G} \,e^{-2\Phi}\Big[-\frac{3}{16}\Omega_{\mu\nu\alpha}\Omega^{\mu\nu\alpha}\Big]\,,
\labell{CS}
\eeqa
and the 6-derivative couplings that result from replacing $H \rightarrow H - (3\alpha'/2) \Omega$ into the 4-derivative action \reef{fourmax} are:
\beqa
{\bf S_2}^{({3})}&=&-\frac{2}{\kappa^2}\int d^{10} x \sqrt{-G} \,e^{-2\Phi}\Big[  \frac{3}{32} H_{\gamma \delta }{}^{\epsilon } R^{\alpha \beta 
\gamma \delta } \Omega _{\alpha \beta \epsilon } -  
\frac{3}{32} F^{\alpha \beta I} F^{\gamma \delta }{}_{I} 
H_{\alpha \gamma }{}^{\epsilon } \Omega _{\beta \delta 
\epsilon }\nn\\&& -  \frac{3}{16} H_{\gamma }{}^{\delta \epsilon } R^{
\alpha \beta }{}_{\alpha }{}^{\gamma } \Omega _{\beta \delta 
\epsilon } + \frac{3}{64} F^{\alpha \beta I} F^{\gamma 
\delta }{}_{I} H_{\alpha \beta }{}^{\epsilon } \Omega 
_{\gamma \delta \epsilon } + \frac{1}{16} H^{\gamma \delta 
\epsilon } R^{\alpha \beta }{}_{\alpha \beta } \Omega 
_{\gamma \delta \epsilon }\nn\\&& -  \frac{3}{16} H_{\beta 
}{}^{\delta \epsilon } R^{\alpha \beta }{}_{\alpha }{}^{\gamma 
} \Omega _{\gamma \delta \epsilon } + \frac{3}{32} H_{\alpha 
\beta }{}^{\epsilon } R^{\alpha \beta \gamma \delta } \Omega 
_{\gamma \delta \epsilon } -  \frac{1}{32} H_{\alpha 
}{}^{\delta \epsilon } H^{\alpha \beta \gamma } H_{\beta 
\delta }{}^{\varepsilon } \Omega _{\gamma \epsilon \varepsilon 
}\nn\\&& + \frac{3}{32} H_{\alpha \beta }{}^{\delta } H^{\alpha \beta 
\gamma } H_{\gamma }{}^{\epsilon \varepsilon } \Omega _{\delta 
\epsilon \varepsilon } -  \frac{1}{192} H_{\alpha \beta \gamma 
} H^{\alpha \beta \gamma } H^{\delta \epsilon \varepsilon } 
\Omega _{\delta \epsilon \varepsilon } + \frac{1}{4} H^{\beta 
\gamma \delta } \Omega _{\beta \gamma \delta } 
\nabla_{\alpha }\nabla^{\alpha }\Phi \nn\\&&-  \frac{1}{4} H^{\beta 
\gamma \delta } \Omega _{\beta \gamma \delta } 
\nabla_{\alpha }\Phi \nabla^{\alpha }\Phi -  \frac{3}{4} 
H_{\alpha }{}^{\gamma \delta } \Omega _{\beta \gamma \delta } 
\nabla^{\beta }\nabla^{\alpha }\Phi \Big]\,.
\labell{CS4}
\eeqa
The total 6-derivative couplings then are:
\beqa
{\bf S_2}&=&{\bf S_2}^{({1})}+{\bf S_2}^{({2})}+{\bf S_2}^{({3})}\,.\labell{S123}
\eeqa
The cosmological symmetry of the action ${\bf S_2}$ when there is no YM field is studied in \cite{Gholian:2023kjj}.

Using the reduction \reef{cRF}, one finds the following reductions for each term in the above equation:
\beqa
\bS_{2}^{c(3)}&\!\!\!\!\!=\!\!\!\!\!&-\frac{2}{\kappa^2}\int dt e^{-\phi}\Bigg[- \frac{1}{128} \dG^{ij} \dG^{kl} \ddot{G}_{ik} \ddot{G}_{jl} + 
\frac{1}{128} \dG_{i}{}^{k} \dG^{ij} \ddot{G}_{j}{}^{l} \ddot{G}_{kl}\Bigg],\labell{eeOO}\\
\bS_{2}^{c(2)}&\!\!\!\!\!=\!\!\!\!\!&-\frac{2}{\kappa^2}\int dt e^{-\phi}\Bigg[- \frac{1}{128} \dcB^{ij} \dG_{i}{}^{k} \dG_{lm} \dG^{lm} \ddot{G}_{jk} 
+ \frac{1}{32} \dcB^{ij} \dG_{i}{}^{k} \dP^2 \ddot{G}_{jk} + 
\frac{1}{32} \dcB^{ij} \dG_{i}{}^{k} \dG_{k}{}^{l} \dP \ddot{G}_{jl} \nn\\&&
+ \frac{1}{64} \dcB^{ij} \dG_{i}{}^{k} \dG_{k}{}^{l} \dG_{l}{}^{m} 
\ddot{G}_{jm} -  \frac{3}{64} \dcB_{i}{}^{k} \dcB^{ij} \dcB_{j}{}^{l} 
\dG_{k}{}^{m} \ddot{G}_{lm} -  \frac{1}{128} \dcB_{ij} \dcB^{ij} 
\dcB^{kl} \dG_{k}{}^{m} \ddot{G}_{lm}\Bigg],\nn\\
\bS_{2}^{c(1)}&\!\!\!\!\!=\!\!\!\!\!&-\frac{2}{\kappa^2}\int dt e^{-\phi}\Bigg[\frac{1}{8} \dA^{iI} \dA^{j}{}_{I} \dA^{kJ} \dA^{l}{}_{J} 
\dcB_{ik} \dcB_{jl} + \frac{1}{16} \dA_{i}{}^{J} \dA^{iI} 
\dA^{j}{}_{I} \dA^{k}{}_{J} \dcB_{j}{}^{l} \dcB_{kl}\nn\\&& + 
\frac{9}{128} \dA_{iI} \dA^{iI} \dA^{jJ} \dA^{k}{}_{J} 
\dcB_{j}{}^{l} \dcB_{kl} -  \frac{1}{128} \dA_{i}{}^{J} \dA^{iI} 
\dA_{jJ} \dA^{j}{}_{I} \dcB_{kl} \dcB^{kl} -  \frac{1}{256} 
\dA_{iI} \dA^{iI} \dA_{jJ} \dA^{jJ} \dcB_{kl} \dcB^{kl}\nn\\&& + 
\frac{1}{32} \dA^{iI} \dA^{j}{}_{I} \dcB_{i}{}^{k} \dcB_{j}{}^{l} 
\dcB_{k}{}^{m} \dcB_{lm} + \frac{7}{512} \dA_{iI} \dA^{iI} 
\dcB_{j}{}^{l} \dcB^{jk} \dcB_{k}{}^{m} \dcB_{lm} + \frac{1}{128} 
\dA^{iI} \dA^{j}{}_{I} \dcB_{i}{}^{k} \dcB_{jk} \dcB_{lm} \dcB^{lm} \nn\\&&
-  \frac{1}{512} \dA_{iI} \dA^{iI} \dcB_{jk} \dcB^{jk} \dcB_{lm} 
\dcB^{lm} + \frac{1}{256} \dcB_{i}{}^{k} \dcB^{ij} \dcB_{j}{}^{l} 
\dcB_{k}{}^{m} \dcB_{l}{}^{n} \dcB_{mn} + \frac{1}{64} \dA^{iI} 
\dA^{j}{}_{I} \dcB_{i}{}^{k} \dcB_{lm} \dcB^{lm} \dG_{jk}\nn\\&& -  
\frac{1}{32} \dA^{iI} \dA^{j}{}_{I} \dA^{kJ} \dA^{l}{}_{J} 
\dG_{ik} \dG_{jl} -  \frac{1}{32} \dA^{iI} \dA^{j}{}_{I} 
\dcB_{i}{}^{k} \dcB_{k}{}^{l} \dcB_{l}{}^{m} \dG_{jm} + 
\frac{1}{128} \dA^{iI} \dA^{j}{}_{I} \dcB_{kl} \dcB^{kl} 
\dG_{i}{}^{m} \dG_{jm} \nn\\&&-  \frac{1}{16} \dA_{i}{}^{J} \dA^{iI} 
\dA^{j}{}_{I} \dA^{k}{}_{J} \dcB_{j}{}^{l} \dG_{kl} + \frac{3}{64} 
\dA_{iI} \dA^{iI} \dA^{jJ} \dA^{k}{}_{J} \dcB_{j}{}^{l} \dG_{kl} + 
\frac{1}{128} \dA^{iI} \dA^{j}{}_{I} \dA^{kJ} \dA^{l}{}_{J} 
\dG_{ij} \dG_{kl} \nn\\&&-  \frac{1}{128} \dA_{iI} \dA^{iI} \dA^{jJ} 
\dA^{k}{}_{J} \dG_{j}{}^{l} \dG_{kl} + \frac{1}{16} \dA^{iI} 
\dA^{j}{}_{I} \dcB_{i}{}^{k} \dcB^{lm} \dG_{jl} \dG_{km} + 
\frac{3}{256} \dA_{iI} \dA^{iI} \dcB^{jk} \dcB^{lm} \dG_{jl} 
\dG_{km} \nn\\&&-  \frac{1}{128} \dcB_{i}{}^{k} \dcB^{ij} \dcB_{l}{}^{n} 
\dcB^{lm} \dG_{jm} \dG_{kn} -  \frac{1}{128} \dA_{i}{}^{J} 
\dA^{iI} \dA_{jJ} \dA^{j}{}_{I} \dG_{kl} \dG^{kl} -  
\frac{1}{256} \dA_{iI} \dA^{iI} \dA_{jJ} \dA^{jJ} \dG_{kl} 
\dG^{kl} \nn\\&&+ \frac{1}{16} \dA^{iI} \dA^{j}{}_{I} \dcB_{i}{}^{k} 
\dcB_{j}{}^{l} \dcB_{k}{}^{m} \dG_{lm} + \frac{1}{16} \dA^{iI} 
\dA^{j}{}_{I} \dcB_{i}{}^{k} \dcB_{k}{}^{l} \dG_{j}{}^{m} \dG_{lm} + 
\frac{1}{32} \dA^{iI} \dA^{j}{}_{I} \dcB^{kl} \dG_{ik} 
\dG_{j}{}^{m} \dG_{lm} \nn\\&&-  \frac{1}{32} \dA^{iI} \dA^{j}{}_{I} 
\dcB_{i}{}^{k} \dcB_{j}{}^{l} \dG_{k}{}^{m} \dG_{lm} -  
\frac{1}{128} \dA_{iI} \dA^{iI} \dcB_{j}{}^{l} \dcB^{jk} 
\dG_{k}{}^{m} \dG_{lm} + \frac{3}{512} \dA_{iI} \dA^{iI} 
\dG_{j}{}^{l} \dG^{jk} \dG_{k}{}^{m} \dG_{lm} \nn\\&&+ \frac{1}{128} 
\dcB_{i}{}^{k} \dcB^{ij} \dcB_{j}{}^{l} \dcB^{mn} \dG_{km} \dG_{ln} + 
\frac{3}{256} \dcB^{ij} \dcB^{kl} \dG_{i}{}^{m} \dG_{j}{}^{n} 
\dG_{km} \dG_{ln} + \frac{1}{128} \dA^{iI} \dA^{j}{}_{I} 
\dcB_{i}{}^{k} \dcB_{jk} \dG_{lm} \dG^{lm} \nn\\&&-  \frac{1}{256} 
\dA_{iI} \dA^{iI} \dcB_{jk} \dcB^{jk} \dG_{lm} \dG^{lm} + 
\frac{1}{64} \dA^{iI} \dA^{j}{}_{I} \dcB_{i}{}^{k} \dG_{jk} 
\dG_{lm} \dG^{lm} + \frac{1}{128} \dA^{iI} \dA^{j}{}_{I} 
\dG_{i}{}^{k} \dG_{jk} \dG_{lm} \dG^{lm} \nn\\&&-  \frac{1}{512} \dA_{iI} 
\dA^{iI} \dG_{jk} \dG^{jk} \dG_{lm} \dG^{lm} -  \frac{1}{32} 
\dcB_{i}{}^{k} \dcB^{ij} \dcB^{lm} \dG_{jl} \dG_{k}{}^{n} \dG_{mn} -  
\frac{1}{128} \dcB_{i}{}^{k} \dcB^{ij} \dcB_{j}{}^{l} 
\dcB_{k}{}^{m} \dG_{l}{}^{n} \dG_{mn} \nn\\&&-  \frac{1}{64} \dcB^{ij} 
\dcB^{kl} \dG_{ik} \dG_{j}{}^{m} \dG_{l}{}^{n} \dG_{mn} + 
\frac{1}{128} \dcB_{i}{}^{k} \dcB^{ij} \dG_{j}{}^{l} \dG_{k}{}^{m} 
\dG_{l}{}^{n} \dG_{mn} -  \frac{5}{32} \dA^{iI} \dA^{j}{}_{I} 
\dA^{kJ} \dcB_{jk} \ddot{A}_{iJ} \nn\\&&+ \frac{3}{32} \dA^{iI} \dA^{j}{}_{I} 
\dA^{kJ} \dG_{jk} \ddot{A}_{iJ} -  \frac{1}{16} \dA^{iI} \dA^{jJ} 
\ddot{A}_{iJ} \ddot{A}_{jI} -  \frac{1}{32} \dA^{iI} \dA^{j}{}_{I} 
\ddot{A}_{i}{}^{J} \ddot{A}_{jJ} -  \frac{1}{16} \dA_{i}{}^{J} \dA^{iI} 
\ddot{A}_{jJ} \ddot{A}^{j}{}_{I} \nn\\&&+ \frac{1}{64} \dA_{iI} \dA^{iI} \ddot{A}_{jJ} 
\ddot{A}^{jJ} -  \frac{1}{32} \dA_{i}{}^{J} \dA^{iI} \dA^{j}{}_{I} 
\dcB_{jk} \ddot{A}^{k}{}_{J} + \frac{3}{64} \dA_{iI} \dA^{iI} \dA^{jJ} 
\dcB_{jk} \ddot{A}^{k}{}_{J} \nn\\&&+ \frac{1}{32} \dA_{i}{}^{J} \dA^{iI} 
\dA^{j}{}_{I} \dG_{jk} \ddot{A}^{k}{}_{J} + \frac{1}{64} \dA_{iI} 
\dA^{iI} \dA^{jJ} \dG_{jk} \ddot{A}^{k}{}_{J} + \frac{1}{32} \dA^{iI} 
\dA^{j}{}_{I} \ddot{\cB}_{i}{}^{k} \ddot{\cB}_{jk} \nn\\&&-  \frac{1}{16} \dA^{iI} 
\dA^{j}{}_{I} \dcB_{i}{}^{k} \dcB_{k}{}^{l} \ddot{\cB}_{jl} -  
\frac{1}{32} \dA^{iI} \dA^{j}{}_{I} \dcB^{kl} \dG_{ik} \ddot{\cB}_{jl} 
-  \frac{1}{32} \dA^{iI} \dA^{j}{}_{I} \dcB_{i}{}^{k} 
\dG_{k}{}^{l} \ddot{\cB}_{jl} \nn\\&&-  \frac{1}{32} \dA^{iI} \dA^{j}{}_{I} 
\dG_{i}{}^{k} \dG_{k}{}^{l} \ddot{\cB}_{jl} -  \frac{1}{128} \dcB^{ij} 
\dcB^{kl} \ddot{\cB}_{ik} \ddot{\cB}_{jl} + \frac{1}{64} \dA_{iI} \dA^{iI} 
\ddot{\cB}_{jk} \ddot{\cB}^{jk} + \frac{1}{32} \dA_{iI} \dA^{iI} \dcB^{jk} 
\dG_{j}{}^{l} \ddot{\cB}_{kl} \nn\\&&+ \frac{1}{128} \dcB_{i}{}^{k} \dcB^{ij} 
\ddot{\cB}_{j}{}^{l} \ddot{\cB}_{kl} + \frac{1}{64} \dcB_{i}{}^{k} \dcB^{ij} 
\dcB^{lm} \dG_{jl} \ddot{\cB}_{km}\! + \!\frac{1}{64} \dcB_{i}{}^{k} 
\dcB^{ij} \dG_{j}{}^{l} \dG_{l}{}^{m} \ddot{\cB}_{km}\! -  \!\frac{1}{64} 
\dcB^{ij} \dcB^{kl} \dG_{ik} \dG_{j}{}^{m} \ddot{\cB}_{lm} \nn\\&&+ \frac{1}{64} 
\dcB_{i}{}^{k} \dcB^{ij} \dcB_{j}{}^{l} \dG_{k}{}^{m} \ddot{\cB}_{lm} + 
\frac{1}{64} \dA_{iI} \dA^{iI} \dA^{jJ} \dA^{k}{}_{J} \ddot{G}_{jk} + 
\frac{1}{16} \dA^{iI} \dA^{j}{}_{I} \ddot{\cB}_{i}{}^{k} \ddot{G}_{jk} + 
\frac{1}{32} \dA^{iI} \dA^{j}{}_{I} \ddot{G}_{i}{}^{k} \ddot{G}_{jk}\nn\\&& -  
\frac{3}{32} \dA^{iI} \dA^{j}{}_{I} \dcB_{i}{}^{k} \dcB_{k}{}^{l} 
\ddot{G}_{jl} -  \frac{1}{32} \dA^{iI} \dA^{j}{}_{I} \dcB^{kl} 
\dG_{ik} \ddot{G}_{jl}\! - \! \frac{1}{32} \dA^{iI} \dA^{j}{}_{I} 
\dcB_{i}{}^{k} \dG_{k}{}^{l} \ddot{G}_{jl}\! - \! \frac{1}{32} \dA^{iI} 
\dA^{j}{}_{I} \dG_{i}{}^{k} \dG_{k}{}^{l} \ddot{G}_{jl}\nn\\&& -  \frac{1}{32} 
\dcB^{ij} \dcB^{kl} \dG_{i}{}^{m} \dG_{km} \ddot{G}_{jl} + \frac{1}{32} 
\dcB^{ij} \dcB^{kl} \ddot{G}_{ik} \ddot{G}_{jl} + \frac{1}{64} \dA_{iI} 
\dA^{iI} \ddot{G}_{jk} \ddot{G}^{jk} + \frac{1}{16} \dA^{iI} \dA^{j}{}_{I} 
\dcB_{i}{}^{k} \dcB_{j}{}^{l} \ddot{G}_{kl}\nn\\&& + \frac{3}{64} \dA_{iI} 
\dA^{iI} \dcB_{j}{}^{l} \dcB^{jk} \ddot{G}_{kl} -  \frac{1}{64} 
\dA_{iI} \dA^{iI} \dG_{j}{}^{l} \dG^{jk} \ddot{G}_{kl} -  \frac{1}{32} 
\dcB_{i}{}^{k} \dcB^{ij} \ddot{\cB}_{j}{}^{l} \ddot{G}_{kl} + \frac{1}{32} 
\dcB_{i}{}^{k} \dcB^{ij} \ddot{G}_{j}{}^{l} \ddot{G}_{kl}\nn\\&& + \frac{1}{32} 
\dcB_{i}{}^{k} \dcB^{ij} \dcB^{lm} \dG_{jl} \ddot{G}_{km} -  
\frac{1}{32} \dcB_{i}{}^{k} \dcB^{ij} \dG_{j}{}^{l} \dG_{l}{}^{m} 
\ddot{G}_{km} + \frac{1}{32} \dcB_{i}{}^{k} \dcB^{ij} \dcB_{j}{}^{l} 
\dcB_{k}{}^{m} \ddot{G}_{lm} \nn\\&&+ \frac{1}{32} \dcB^{ij} \dcB^{kl} \dG_{ik} 
\dG_{j}{}^{m} \ddot{G}_{lm} -  \frac{1}{32} \dcB_{i}{}^{k} \dcB^{ij} 
\dcB_{j}{}^{l} \dG_{k}{}^{m} \ddot{G}_{lm}\Bigg].\nn
\eeqa
Obviously, the above one-dimensional actions are not invariant under the $O(9,25)$-transformations. We have to add total derivative terms and perform field redefinitions to make them invariant.

We add the following total derivative term at order $\alpha'^2$ to $\!\!\bS_2^c$:
\beqa
\cJ_2&=&-\frac{2}{\kappa^2}\int dt\frac{d}{dt}(e^{-\phi}\cI_2)\,,
\eeqa
where $\cI_2$ includes all possible terms at the five-derivative level, which are constructed from $\dP$, $\dB$, $\dG$, $\dA$, $\ddot{\phi}$, $\ddot{B}$, $\ddot{G}$, $\ddot{A}$, $\cdots$,  with arbitrary coefficients.

We also make the following field redefinitions:
\begin{eqnarray}
G_{ij}&\rightarrow &G_{ij}+\alpha' \delta G^{(1)}_{ij}+\frac{1}{2}\alpha'^2 \delta G^{(2)}_{ij},\nn\\
B_{ij}&\rightarrow &B_{ij}+ \alpha'\delta B^{(1)}_{ij}+\frac{1}{2} \alpha'^2\delta B^{(2)}_{ij},\nn\\
A_{iI}&\rightarrow &A_{iI}+ \alpha'\delta A^{(1)}_{iI}+\frac{1}{2} \alpha'^2\delta A^{(2)}_{iI},\nn\\
\phi &\rightarrow &\phi+ \alpha'\delta\phi^{(1)}+\frac{1}{2} \alpha'^2\delta\phi^{(2)},\nn\\
n &\rightarrow &n+ \alpha' \delta n^{(1)}+\frac{1}{2} \alpha'^2 \delta n^{(2)},\labell{gbpn2}
\end{eqnarray}
where the first-order perturbations $\delta G^{(1)}_{ij}$, $\delta B^{(1)}_{ij}$, $\delta A^{(1)}_{iI}$, $\delta \phi^{(1)}$, $\delta n^{(1)}$ are given in \reef{dG1dB1}, and the second-order perturbations $\delta G^{(2)}_{ij}$, $\delta B^{(2)}_{ij}$, $\delta A^{(2)}_{iI}$, $\delta \phi^{(2)}$, $\delta n^{(2)}$ consist of all possible terms at the four-derivative level constructed from $\dP$, $\dB$, $\dG$, $\dA$, $\ddot{\phi}$, $\ddot{B}$, $\ddot{G}$, $\ddot{A}$, $\cdots$, with arbitrary coefficients. Using the relation \reef{tcB}, one finds the perturbation $\delta \dcB^{(2)}_{ij}$ is related to the perturbations $\delta B^{(2)}_{ij}$ and $\delta A^{(2)}_{iI}$ as $\delta \dcB^{(2)}_{ij} = \frac{d}{dt}(\delta B_{ij}^{(2)}) - (\dA_{iI}\delta A^{(2)}_j{}^{I} - \dA_{jI}\delta A^{(2)}_i{}^{I})$.

When the field variables in $S_0^c$ are changed according to the above field redefinitions, they produce two sets of couplings at order $\alpha'^2$. One set is produced by the second-order perturbations at order $\alpha'^2$ which is given by $\frac{1}{2}\delta_2 S_0^c$. This term is the same as \reef{dS0c} in which the first-order perturbation $\alpha'\delta\psi^{(1)}$ is replaced by the second-order perturbation $\frac{1}{2}\alpha'^2\delta\psi^{(2)}$.
The other set is reproduced by two first-order perturbations. Since $\delta A^{(1)}_{iI}$ is zero, this set is
\beqa
\frac{1}{2}\delta_1^2 S_0^c&=&-\frac{2}{\kappa^2}\int dt e^{-\Phi}\Bigg[\frac{1}{4} \frac{d}{dt}\delta B^{(1)}_{ij} \frac{d}{dt}\delta B^{(1)}{}^{ij} + \frac{1}{4} 
\frac{d}{dt}\delta G^{(1)}_{ij} \frac{d}{dt}\delta G^{(1)}{}^{ij} -  \frac{d}{dt}\delta \phi^{(1)} \frac{d}{dt}\delta \phi^{(1)}\nn\\&& -  \dcB^{ij} \frac{d}{dt}\delta B^{(1)}_{i}{}^{k} \delta G^{(1)}_{jk} -  \dG^{ij} \frac{d}{dt}\delta G^{(1)}_{i}{}^{k} \delta G^{(1)}_{jk} + \Big(\frac{1}{4} \dcB^{ij} \dcB^{kl} + \frac{1}{4} \dG^{ij} \dG^{kl}\Big)
\delta G^{(1)}_{ik} \delta G^{(1)}_{jl}\nn\\&& + \Big(\frac{1}{2}\dA^{kI}\dA^j{}_I+\frac{1}{2} \dcB_{i}{}^{k} 
\dcB^{ij}+ \frac{1}{2} 
\dG_{i}{}^{k} \dG^{ij} \Big) \delta G^{(1)}_{j}{}^{l} \delta G^{(1)}_{kl} -  
\frac{1}{2} \dcB^{ij} \frac{d}{dt}\delta B^{(1)}_{ij} \delta n^{(1)}\nn\\&& -  \frac{1}{2} \dG^{ij} \frac{d}{dt}\delta G^{(1)}_{ij} \delta n^{(1)} + 2 \dP \frac{d}{dt}\delta \phi^{(1)} 
\delta n^{(1)} +\Big(\frac{1}{2}\dA^{kI}\dA^j{}_I+ \frac{1}{2} \dcB_{i}{}^{k} \dcB^{ij}\! + \!\frac{1}{2} \dG_{i}{}^{k} \dG^{ij}\Big)\delta G^{(1)}_{jk} 
\delta n^{(1)}  \nn\\&& +\Big( \frac{1}{2}\dA_{iI}\dA^{iI}+\frac{1}{4} \dcB_{ij} \dcB^{ij} + 
\frac{1}{4} \dG_{ij} \dG^{ij}  -  \dP^2\Big) 
\delta n^{(1)}\delta n^{(1)}\nn\\&& + \Big( \frac{1}{4}\dA_{iI}\dA^{iI}+\frac{1}{8} \dcB_{ij} \dcB^{ij}  + \frac{1}{8} \dG_{ij} \dG^{ij} -  
\frac{1}{2} \dP^2\Big) \delta\phi^{(1)} \delta \phi^{(1)} \nn\\&& -  \frac{1}{2} \dG^{ij} \frac{d}{dt}\delta G^{(1)}_{ij} \delta \phi^{(1)} + 2 \dP \frac{d}{dt}\delta \phi^{(1)} \delta 
\phi^{(1)} +\Big(\frac{1}{2}\dA^{kI}\dA^j{}_I+ \frac{1}{2} \dcB_{i}{}^{k} \dcB^{ij}+ \frac{1}{2} \dG_{i}{}^{k} \dG^{ij} \Big)\delta G^{(1)}_{jk} 
\delta \phi^{(1)} \nn\\&& -  \frac{1}{2} 
\dcB^{ij} \frac{d}{dt}\delta B^{(1)}_{ij} \delta \phi^{(1)}+\Big( \frac{1}{2}\dA_{iI}\dA^{iI}+ \frac{1}{4} \dcB_{ij} \dcB^{ij} + \frac{1}{4} \dG_{ij} \dG^{ij} -  \dP^2\Big) \delta n^{(1)} \delta \phi^{(1)} \Bigg]\,,\labell{d1}
\eeqa
where the first-order perturbations are given in  \reef{dG1dB1}.

When the field variables in the $O(9,25)$-invariant action $S_1^c$ at order $\alpha'$, as given by  \reef{action2}, are changed according to the field redefinition in  \reef{gbpn2}, one also finds the couplings at order $\alpha'^2$ that result from the first-order perturbation. Since $\delta A^{(1)}_{iI}$ is zero, they are
\beqa
\delta_1 S_1^c&=&-\frac{2}{8\kappa^2}\int dt e^{-\Phi}\Bigg[- \dcB^{ij} \dG_{i}{}^{k} \dG_{k}{}^{l} \frac{d}{dt}\delta B^{(1)}_{jl} + 
\frac{1}{2} \dcB_{i}{}^{k} \dcB^{ij} \dcB_{j}{}^{l} \frac{d}{dt}\delta B^{(1)}_{kl} - 
 \frac{1}{2} \dcB^{ij} \dG_{i}{}^{k} \dG_{j}{}^{l} \frac{d}{dt}\delta B^{(1)}_{kl} \nn\\&&
-  \frac{1}{2} \dcB^{ij} \dcB^{kl} \dG_{ik} \frac{d}{dt}\delta G^{(1)}_{jl} + 
\dcB_{i}{}^{k} \dcB^{ij} \dG_{j}{}^{l} \frac{d}{dt}\delta G^{(1)}_{kl} + \frac{1}{2} 
\dG_{i}{}^{k} \dG^{ij} \dG_{j}{}^{l} \frac{d}{dt}\delta G^{(1)}_{kl}\nn\\&& -  
\frac{1}{2} \dcB^{ij} \dcB^{kl} \dG_{i}{}^{m} \dG_{km} \delta 
G^{(1)}_{jl} -  \dcB_{i}{}^{k} \dcB^{ij} \dG_{j}{}^{l} \dG_{l}{}^{m} \delta 
G^{(1)}_{km} -  \frac{1}{2} \dcB_{i}{}^{k} \dcB^{ij} \dcB_{j}{}^{l} 
\dcB_{k}{}^{m} \delta G^{(1)}_{lm} \nn\\&&+ \dcB^{ij} \dcB^{kl} \dG_{ik} 
\dG_{j}{}^{m} \delta G^{(1)}_{lm} -  \frac{1}{2} \dcB_{i}{}^{k} \dcB^{ij} 
\dG_{j}{}^{l} \dG_{k}{}^{m} \delta G^{(1)}_{lm} -  \frac{1}{2} 
\dG_{i}{}^{k} \dG^{ij} \dG_{j}{}^{l} \dG_{k}{}^{m} \delta G^{(1)}_{lm}\nn\\&& -  
\frac{3}{8} \dcB_{i}{}^{k} \dcB^{ij} \dcB_{j}{}^{l} \dcB_{kl} \delta 
n^{(1)} + \frac{3}{4} \dcB^{ij} \dcB^{kl} \dG_{ik} \dG_{jl} \delta n^{(1)} -  
\frac{3}{2} \dcB_{i}{}^{k} \dcB^{ij} \dG_{j}{}^{l} \dG_{kl} \delta 
n^{(1)}\nn\\&& -  \frac{3}{8} \dG_{i}{}^{k} \dG^{ij} \dG_{j}{}^{l} \dG_{kl} 
\delta n^{(1)} -  \frac{1}{8} \dcB_{i}{}^{k} \dcB^{ij} \dcB_{j}{}^{l} 
\dcB_{kl} \delta \phi^{(1)} + \frac{1}{4} \dcB^{ij} \dcB^{kl} \dG_{ik} 
\dG_{jl} \delta \phi^{(1)}\nn\\&& -  \frac{1}{2} \dcB_{i}{}^{k} \dcB^{ij} 
\dG_{j}{}^{l} \dG_{kl} \delta \phi^{(1)} -  \frac{1}{8} \dG_{i}{}^{k} 
\dG^{ij} \dG_{j}{}^{l} \dG_{kl} \delta \phi^{(1)}
- \dA_{i}{}^{J} \dA^{iI} \dA^{j}{}_{I} \dA^{k}{}_{J} \delta 
G^{(1)}{}_{jk} \nn\\&& + \dA^{iI} \dA^{j}{}_{I} \dcB_{i}{}^{k} \dcB_{k}{}^{l} 
\delta G^{(1)}{}_{jl}+ \dA^{iI} \dA^{j}{}_{I} \dcB^{kl} \dG_{ik} 
\delta G^{(1)}{}_{jl} -  \dA^{iI} \dA^{j}{}_{I} \dcB_{i}{}^{k} 
\dG_{k}{}^{l} \delta G^{(1)}{}_{jl}  \nn\\&&-  \dA^{iI} \dA^{j}{}_{I} 
\dG_{i}{}^{k} \dG_{k}{}^{l} \delta G^{(1)}{}_{jl} -  \frac{1}{2} 
\dA^{iI} \dA^{j}{}_{I} \dcB_{i}{}^{k} \dcB_{j}{}^{l} \delta 
G^{(1)}{}_{kl} -  \dA^{iI} \dA^{j}{}_{I} \dcB_{i}{}^{k} \dG_{j}{}^{l} 
\delta G^{(1)}{}_{kl}  \nn\\&&-  \frac{1}{2} \dA^{iI} \dA^{j}{}_{I} 
\dG_{i}{}^{k} \dG_{j}{}^{l} \delta G^{(1)}{}_{kl} -  \frac{3}{2} 
\dA_{i}{}^{J} \dA^{iI} \dA_{jJ} \dA^{j}{}_{I} \delta n^{(1)} -  
\frac{3}{2} \dA^{iI} \dA^{j}{}_{I} \dcB_{i}{}^{k} \dcB_{jk} 
\delta n^{(1)}  \nn\\&&- 3 \dA^{iI} \dA^{j}{}_{I} \dcB_{i}{}^{k} \dG_{jk} 
\delta n^{(1)} -  \frac{3}{2} \dA^{iI} \dA^{j}{}_{I} 
\dG_{i}{}^{k} \dG_{jk} \delta n^{(1)} -  \frac{1}{2} \dA_{i}{}^{J} 
\dA^{iI} \dA_{jJ} \dA^{j}{}_{I} \delta \phi^{(1)}\nn\\&& -  \frac{1}{2} 
\dA^{iI} \dA^{j}{}_{I} \dcB_{i}{}^{k} \dcB_{jk} \delta \phi^{(1)} - 
 \dA^{iI} \dA^{j}{}_{I} \dcB_{i}{}^{k} \dG_{jk} \delta \phi^{(1)} - 
 \frac{1}{2} \dA^{iI} \dA^{j}{}_{I} \dG_{i}{}^{k} \dG_{jk} 
\delta \phi^{(1)} \nn\\&&+ \dA^{iI} \dA^{j}{}_{I} \dcB_{i}{}^{k} 
\frac{d}{dt}(\delta G^{(1)}{}_{jk}+\delta B^{(1)}{}_{jk}) + \dA^{iI} 
\dA^{j}{}_{I} \dG_{i}{}^{k} \frac{d}{dt}(\delta G^{(1)}{}_{jk}+\delta B^{(1)}{}_{jk})
\Bigg]\,,\labell{d2}
\eeqa
where we have inserted the lapse function in the $O(9,25)$-invariant action \reef{action2} by replacing $dt$ with $dt/n^3$.

By inserting the first-order perturbations  \reef{dG1dB1} into \reef{d1}, \reef{d2}, and inserting the arbitrary second-order perturbations into $\frac{1}{2}\delta_2S_0^c$, one finds that the cosmological action \reef{S0122}  can be written as:
\beqa
S_2^c=\!\!\bS_2^c+\delta_1S_1^c-\frac{1}{2}\delta_1^2S_0^c+\frac{1}{2}\delta_2S_0^c+\cJ_2=0\,,\labell{action21}
\eeqa
for some specific values of the parameters in field redefinitions and total derivative terms. Since we are not interested in studying the couplings at order $\alpha'^3$, we don't write the explicit form of the second-order field redefinitions here. The above result indicates that the coefficient $c_{3,0}$ in \reef{cosmH} is zero, which is the same result that has been found in \cite{Codina:2021cxh} by studying only the pure gravity part.

It has been found in \cite{Garousi:2024imy} that the truncated T-duality transformations produce 469 relations between the six-derivative couplings in the maximal basis. This constraint has also been imposed in \cite{Garousi:2024imy} on a basis which has 468 couplings, and it was found that the T-duality produces another set of six-derivative couplings which involve 107 terms and are not in the canonical form as in \reef{L22}. We have done the same cosmological calculation for this expression for the six-derivative couplings and found exactly the same result. This ends our illustration that the cosmological reduction of the couplings found in \cite{Garousi:2024imy} by the truncated T-duality transformation is fully consistent with the expected $O(9,25)$-symmetry.

\newpage

\section{Discussion}

In this paper, we have proposed that the cosmological reduction of the bosonic part of the classical effective action of the heterotic string theory at any order of $\alpha'$ when the YM field is restricted to the Cartan subalgebra of the $SO(32)$ or $E_8\times E_8$ group can be written in a canonical form. This canonical form includes only the first time-derivative of the $O(9,25)$-valued generalized metric $\mathcal{H}$, which encompasses the NS-NS and YM fields \cite{Maharana:1992my,Hohm:2014sxa}. This proposal, denoted as \reef{cosmH}, is an extension of the canonical form that has been proposed by Hohm and Zwiebach for only the NS-NS couplings \reef{cosm}, where the $O(9,9)$-valued generalized metric $\mathcal{S}$ is extended to $\mathcal{H}$.

We then show that the cosmological reductions of the 4- and 6-derivative couplings that have been recently found by the truncated T-duality transformations on the circular reduction of the heterotic couplings \cite{Garousi:2024avb,Garousi:2024imy} exactly satisfy the canonical form \reef{cosmH}. This canonical form does not have any trace of an odd number of $\dot{\mathcal{H}}$, which also confirms the result found in \cite{Garousi:2024vbz} that the effective action of heterotic string theory does not have any odd-derivative couplings.

The consistency demonstrated above highlights the power of the truncated T-duality approach in deriving the higher-derivative effective actions of heterotic string theory. In applying this technique, one uses the fact that the coupling constants are independent of the YM gauge. Hence, the calculation can be performed in a particular gauge in which the YM field is zero, but its derivatives are not. In this gauge, the commutator terms in the YM field strength and its derivatives become zero.
It is important to note that we do not use the Cartan subalgebra in which the YM fields commute. We only use a particular YM gauge. After finding the coupling constants in this particular gauge, the result can then be extended to any other gauges in which the commutator terms are not zero. Hence, the result from the truncated T-duality approach is valid for YM gauge fields belonging to the full algebra of the $SO(32)$ or $E_8\times E_8$ groups in the heterotic string theories.

The YM gauge field can also be extended to the $O(16)\times O(16)$ group of the bosonic tachyon-free heterotic string theory \cite{Alvarez-Gaume:1986ghj,Dixon:1986iz}. In both the supersymmetric and bosonic heterotic string theories, there are fermions that should be included in the T-duality transformation. The fermionic T-duality transformations have been studied in the type IIB superstring theory \cite{Berkovits:2008ic,Beisert:2008iq}. It would be interesting to find the fermionic T-duality transformation in the heterotic string theory and include it in the truncated T-duality transformation. This would allow for the investigation of fermionic couplings, as well as the coupling of the NS-NS and YM fields, and their inclusion in the generalized metric $\mathcal{H}$ to study their cosmological symmetry.

 

\end{document}